\documentclass[fleqn,usenatbib]{mnras}

\usepackage{newtxtext,newtxmath}

\usepackage[T1]{fontenc}
\usepackage{ae,aecompl}

\usepackage{graphicx}	
\usepackage{amsmath}	
\usepackage{amssymb}	
\usepackage[alsoload=parsec]{siunitx}
\usepackage{physics}
\usepackage{hyperref}

\usepackage{bm}
\usepackage{xcolor}

\newcommand{\chieff}{{\chi_{\rm eff}}}
\newcommand\hl[1]{%
  \bgroup
  \hskip0pt\color{red!80!black}%
  #1%
  \egroup
}

\graphicspath{{figures/}}

\title[Constraints on BBH populations]{Constraints on Binary Black Hole Populations from LIGO--Virgo Detections}

\author[J. Roulet and M. Zaldarriaga]{
Javier Roulet,$^{1}$\thanks{E-mail: jroulet@princeton.edu (JR)}
Matias Zaldarriaga,$^{2}$
\\
$^{1}$Department of Physics, Princeton University, Princeton, NJ, 08540, US\\
$^{2}$School of Natural Sciences, Institute for Advanced Study, Princeton, NJ, 08540, US\\
}

\date{Accepted XXX. Received YYY; in original form ZZZ}

\pubyear{2018}

\begin{document}
\label{firstpage}
\pagerange{\pageref{firstpage}--\pageref{lastpage}}
\maketitle

\begin{abstract}
We reanalyse the LIGO--Virgo strain data of the 10 binary black hole mergers reported to date and compute the likelihood function in terms of chirp mass, mass ratio and effective spin. We discuss the strong degeneracy between mass ratio and spin for the three lighter events. We use this likelihood and an estimate of the horizon volume as a function of intrinsic parameters to constrain the properties of the population of merging binary black holes. The data disfavour large spins. Typical spins are constrained to $\overline a \lesssim 0.4$, even if the underlying population has randomly-oriented spins. For aligned spins the constraints are tighter, with typical spins required to be around $\overline a\sim 0.1$ and have comparable dispersion. We detect no statistically significant tendency towards a positive average spin in the direction of the orbital angular momentum. We put an upper limit on the fraction of systems where the secondary could have been tidally locked prior to the formation of the black holes (corresponding to merger times shorter than $10^8$ years) $f \lesssim 0.3$. 
Four events are consistent with having a maximally-spinning secondary, although one only marginally.
We confirm previous findings that there is a hint of a cutoff at high mass. The data favour distributions of mass ratios with an average $\overline q \gtrsim 0.7$.
\end{abstract}

\begin{keywords}
gravitational waves -- stars: black holes -- methods: data analysis
\end{keywords}



\section{Introduction}

Binary black holes (BBH) have been observed for the first time with the recent advent of gravitational wave (GW) observatories \citep{GW150914,GW151226,O1,GW170104,GW170608,GW170814,O2catalog}. The astrophysical origin of these systems remains a major open question. Potential formation channels that have been proposed in the literature include isolated binary evolution through a common envelope phase, a chemically homogeneous evolution in a tidally locked
binary, the dynamical formation in dense stellar environments such as globular clusters or in triple systems and the formation in galactic nuclear disks assisted by the presence of gas (references to various scenarios and how they fare in comparison with the LIGO data can be found in \citet{O1}). Although the gravitational wave data are yet insufficient to decide between these scenarios, constraints on specific models can already start to be set (see for example results in \citet{O1,Vitale2017,Talbot2017,Farr2017,Hotokezaka:2017dun,Farr2018,O2populations}).

To use the LIGO--Virgo events to constrain the properties of the population of merging black holes we need the likelihood of individual events as a function of the parameters of the binary. Although constraints on individual parameters have been reported by the LIGO and Virgo collaborations (LVC), the full likelihoods have not been provided to the community yet (posterior samples have been recently released). These are necessary in order to properly include the correlations between parameters, which as we will see are substantial. Thus in this paper we set out to reanalyse the LIGO--Virgo strain data (under some simplifying assumptions) to obtain likelihoods for the events and then use those to set constraints on the parameters of population models for the BBHs. The attempts to put constraints on the populations already presented in the literature were limited by the lack of the individual event likelihoods, so approximate constraints were based on properties of the one-dimensional posteriors gathered from the LVC figures and papers. 

Neglecting center-of-mass acceleration (e.g. due to a third body), BBH merger events are completely characterized by 17 parameters (two masses, six spin components, two angles for the orbit orientation, two for the sky location, distance, merger phase and time, orbital eccentricity and pericenter angle).
However, the following key observations allow us to significantly reduce the number of parameters considered. 

First, not all the parameters are informative of the population properties. Population models generally predict a homogeneous and isotropic GW-source distribution on scales resolvable by GW detectors, so the source location is irrelevant to distinguish among these models. Similarly, the source orientation and the time and phase of merger are uniformly distributed. Moreover, the signal dependence on these parameters is well understood, so the likelihood can be marginalized over them.

Second, not all the parameters can be constrained by the data at the current sensitivity levels. Only one of the spin combinations, $\chieff$, is relatively well measured (see for example \citet{Vitale2014,Vitale2017_par_est}). It is defined by
\begin{equation} \label{eq:chi_eff}
\chieff = \frac{m_1 \vec a_1 + m_2 \vec a_2}{m_1 + m_2} \cdot \hat L,
\end{equation}
\noindent where $\vec a_i = c \vec S_i / G m_i^2$ and $\hat L$ is the direction of the orbital angular momentum. Spin components orthogonal to $\chieff$ are largely unconstrained. Thus, including these parameters in the analysis increases the computational cost without significantly changing the results. Moreover, the waveform templates currently used for detection and parameter estimation have no eccentricity, so LIGO is not able to measure it. We will assume no eccentricity as well.

Third, astrophysical models for the populations are quite crude and thus small shifts in the parameters or error bars are not likely to change the astrophysical conclusions one might draw. At this stage one is interested in more qualitative questions such as whether the BHs are spinning fast, whether the spins tend to be aligned normal to the orbital plane or what the range of masses of the BBHs is.

As a result of these considerations, there is a clear hierarchy in the parameters based on how much they can constrain BBH population models. $\chieff$ and the two masses are measurable and have distributions dependent on the models \citep{Mandel2010,Rodriguez2016,Zevin2017,Stevenson2017}. The other spin components and the eccentricity, while dependent on the population model, are still poorly constrained by the data, so we henceforth ignore them. The various angles, phase, time of merger and distance are uninformative of the population, so they are nuisance parameters for our purposes. They have a known effect on the signal, so we marginalize over them in \S\ref{sec:likelihood}.

We will parametrize the two masses in terms of the chirp mass $\mathcal M$ and the mass ratio $q$, given by
\begin{equation} \label{eq:Mchirp_q}
	\begin{split}
		\mathcal M &= \frac{(m_1 m_2)^{3/5}}{(m_1 + m_2)^{1/5}} \\
		q &= \frac{m_2}{m_1} \leq 1,
	\end{split}
\end{equation}
because $\mathcal M, q$ are generally less correlated than $m_1, m_2$, as these are the combinations that respectively enter the GW waveform at the leading and the next-to-leading order in the post-Newtonian expansion. To implement our restriction to a single spin variable, we will use the prescription that the spins are aligned with the orbital angular momentum, and that $\chi_1 = \chi_2 = \chieff$, where $\chi_i = \vec a_i \cdot \hat L$. At the current level of sensitivity the exact way one distributes the spin to obtain a given $\chieff$ does not affect the parameter constraints in a meaningful way.

We will denote the informative parameters by $\mathbf{p} = (\mathcal M, q, \chieff)$. Unless otherwise stated, we will refer to the detector-frame mass, whose value is redshifted from the source-frame mass by $\mathcal M = (1+z) \mathcal M_{\rm source}$. Both $q$ and $\chieff$ are independent of redshift.

\section{Single-event likelihood}
\label{sec:likelihood}

In this section we describe our computation of the BBH-parameter likelihood of a GW event and its analytical marginalization over the nuisance parameters. What we do is rather standard but we detail our procedure so that we can report all our simplifications. We first analyse the case of a single detector.

We will define the noise-weighted inner product between two functions in the frequency domain
\begin{equation} \label{eq:product}
	\langle x \mid y \rangle = 4 \Re \int_0^\infty \frac{\tilde x^*(f) \tilde y(f)}{\sigma^2(f)} {\rm d}f,
\end{equation}
where $\sigma^2(f)$ is the one-sided power spectral density (PSD) of the detector noise and the tildes indicate Fourier transforms.
Under the assumption that the noise is additive, stationary and Gaussian, the single-detector likelihood $\mathcal L = P(d \mid h)$ that the data $d$ have been produced by a model GW signal $h$ is
\begin{equation} \label{eq:logL}
	\log\mathcal L = -\frac{1}{2} \langle d-h \mid d-h \rangle.
\end{equation}
We have access to both $d$ and $h$, because the strain data for all reported events and the approximants for generating template waveforms have been released by the Gravitational Wave Open Science Center (GWOSC) \citep{Vallisneri2015}. We use two different approximants, \texttt{SEOBNRv4\_ROM}, based on the effective-one-body formalism \citep{Bohe2017}, and \texttt{IMRPhenomD}, based on a phenomenological approach \citep{Khan2016}, as a robustness test of our results. We estimate the PSD using the \texttt{PyCBC} \citep{Biwer2018} implementation of the median-mean spectrum described in \citet{Allen2012} on a segment of $\SI{32}{\second}$ of data centered around each event. As recommended by the GWOSC, we use a lower frequency cutoff of $\SI{10}{\hertz}$ except for GW170608, which we cut at $\SI{20}{\hertz}$ for Livingston and $\SI{30}{\hertz}$ for Hanford.

We decompose the modeled signal into the form
\begin{equation} \label{eq:h}
	\tilde h(f; \mathbf{p}, a_0, t_0, \phi_0) = a_0 {\rm e}^{{\rm i}(2 \pi f t_0 - \phi_0)} \frac{\tilde h_0(f;\mathbf{p})}{\sqrt{\langle h_0 \mid h_0 \rangle}}.
\end{equation}
Here $\tilde h_0 (f; \mathbf{p})$ is the waveform template, that depends on the set of physical parameters $\mathbf{p} = (\mathcal M, q, \chieff)$ that are intrinsic to the binary. 
We work in the quadrupole radiation approximation, i.e. that the dominant spherical harmonics of the emitted GW are $(l, m) = (2,\pm2)$. The shape of $h_0$ is then independent of the nuisance parameters, which only enter through $a_0, \phi_0, t_0$. The inclusion of other multipole moments would introduce a dependence of the waveform shape on the orientation of the binary \citep{Cotesta2018}. 
We divide $h_0$ by the norm to eliminate its arbitrary normalization. We may then compute $h_0$ at any fiducial configuration, e.g. directly above the detector at a $\SI{1}{\mega\parsec}$ distance, with a face-on alignment.
$t_0$ and $\phi_0$ are the time and phase of the merger as seen in the detector. $a_0$ is the amplitude, which can be interpreted as the expectation value for the signal-to-noise ratio (SNR) with which a signal identical to $h$ would be measured in the detector, given by $\sqrt{\langle h \mid h \rangle}$.
For multiple detectors, each one will have different values of $a_0, t_0, \phi_0$ that are related by the various angles between source and detectors.

Combining \eqref{eq:logL} and \eqref{eq:h} we obtain
\begin{equation} \label{eq:logL2}
	\begin{aligned} 
		\log\mathcal L &= -\frac{1}{2} \left(\langle d \mid d \rangle
						  - 2 \langle d \mid h \rangle
						  + \langle h \mid h \rangle \right) \\
					   &= -\frac{1}{2} \left(\langle d \mid d \rangle
						  - 2 a_0 \abs{z} \cos(\phi_0 - \arg z)
						  + a_0^2 \right),			 
	\end{aligned}
\end{equation}
\noindent where $z({\mathbf p}, t_0)$ is the complex matched filter output \citep{Allen2012}:
\begin{equation} \label{eq:z}
	z({\mathbf p}, t_0) = \frac{4}{\sqrt{\langle h_0 \mid h_0 \rangle}} \int_0^\infty \frac{\tilde d^*(f) \tilde h_0(f; {\mathbf p})}{\sigma^2(f)} {\rm e}^{{\rm i} 2\pi f t_0} {\rm d}f.
\end{equation}
Note that, for any set of parameters $\mathbf{p}$, we can compute $z$ for all $t_0$ with a single Fast Fourier Transform.

\subsection{Likelihood marginalization}

We will now marginalize the likelihood \eqref{eq:logL2} over the nuisance parameters $a_0, \phi_0, t_0$, since their priors do not depend on the BBH population model, and we will keep the dependence on $\mathbf{p}$.
Our approach will be to do the marginalization subject to the condition $\mathcal D$ that the event has been detected, that is, we will define
\begin{align} \label{eq:Lmarg_def}
    \overline{\mathcal L}(\mathbf{p}) &\equiv P(d \mid \mathbf{p}, \mathcal D) \notag\\
    &= \int {\rm d}a_0 {\rm d}\phi_0 {\rm d}t_0 
      P_{\rm prior}(a_0, \phi_0, t_0 \mid \mathbf{p}, \mathcal D)
      \mathcal L(a_0, \phi_0, t_0, \mathbf{p} \mid \mathcal D) \notag\\
    &= \int {\rm d}a_0 {\rm d}\phi_0 {\rm d}t_0 
      P_{\rm prior}(a_0, \phi_0, t_0 \mid \mathcal D)
      \mathcal L(a_0, \phi_0, t_0, \mathbf{p}).
\end{align}
The last equality follows because, as we will show, the prior for $a_0, \phi_0, t_0$ does not depend on $\mathbf{p}$ once conditioned to detectability; and the likelihood $\mathcal L$ is suppressed for combinations of parameters that yield undetectable signals, since the data under consideration correspond to detections and detectability is a property of the data only.
$P(d \mid \mathbf{p}, \mathcal D)$ differs from $P(d \mid \mathbf{p})$ in that it excludes the selection bias of the detector, whose sensitivity depends on $\mathbf{p}$.
Indeed, for data corresponding to detections
\begin{equation} \label{eq:P(d|p)}
  \begin{split}
      P(d \mid \mathbf{p})
      &= P(d \mid \mathcal D, \mathbf{p}) P(\mathcal D \mid \mathbf{p}),
  \end{split}
\end{equation}
where $P(\mathcal D \mid \mathbf{p})$ is the selection bias (see Appendix~\ref{app:bias}). By imposing the detection condition $\mathcal D$ we are taking the observer's point of view, where the events analysed are conditioned to detection and the selection bias enters in the form of a prior for $\mathbf{p}$. We show the equivalence to the alternative, physical approach usually taken in the literature (e.g. \citet{Fishbach:2017zga, Mandel2018}) in Appendix~\ref{app:bias}.

We compute the prior in Eq.~\eqref{eq:Lmarg_def} as follows.
$\phi_0$ and $t_0$ have uniform priors. $a_0$ is proportional to
$D^{-1}$, where $D$ is the luminosity distance to the event. At low redshifts, the prior for the distance is $P_{\rm prior}(D) \propto D^2$, so demanding $P_{\rm prior}(D) {\rm d}D = P_{\rm prior}(a_0) {\rm d}a_0$ yields 
$P_{\rm prior}(a_0) \propto a_0^{-4}$.
For the event to have been a detection, $a_0$ must exceed a certain threshold value, for which we take a conservative (low) value of $a_{0,{\rm min}} = 9$ (see fig.~9 of \citet{O1}).

The priors for the nuisance parameters are then
\begin{align}
	P_{\rm prior}(\phi_0) &= \frac{1}{2\pi} \label{eq:phi0_prior} \\
	P_{\rm prior}(t_0) &= \frac{1}{T} \label{eq:t0_prior} \\
	P_{\rm prior}(a_0) &= \frac{3 a_{0,{\rm min}}^3}{a_0^4}, 
    	\qquad a_0 > a_{0,{\rm min}} \label{eq:a0_prior}
\end{align}
where $T$ is the duration of the data being analysed and the priors have been normalized to integrate to $1$ over their domains, as required by the detection condition.

Using \eqref{eq:logL2}, \eqref{eq:phi0_prior} and \eqref{eq:a0_prior}, the marginalization of the likelihood over $\phi_0$ and $a_0$ yields:
\begin{equation} \label{eq:Lmarg_a0_phi0}
 	\int_{a_{0,{\rm min}}}^\infty {\rm d}a_0 P_{\rm prior}(a_0)
 		\int_0^{2 \pi} \frac{{\rm d}\phi_0}{2\pi} \mathcal L
 		= 3 a_{0,{\rm min}}^3 {\rm e}^{-\frac{1}{2}\langle d \mid d \rangle} I(\abs{z}),
\end{equation} 
where we have defined
\begin{equation}
 	I(\abs{z}) = \int_{a_{0,{\rm min}}}^\infty
 		\frac{I_0(a_0 \abs{z}) {\rm e}^{-\frac{1}{2}a_0^2}}{a_0^4} {\rm d}a_0.
\end{equation}
Here, $I_0$ is the modified Bessel function of the first kind of order zero. In practice, we ignore the constant factor in front of $I(\abs{z})$ in \eqref{eq:Lmarg_a0_phi0} as it does not depend on the parameters. To implement the computation of $I(\abs{z})$ efficiently, we tabulate its more smoothly varying logarithm for several values of $\abs{z}$ and interpolate in between.

Finally, we can further marginalize Eq.~\eqref{eq:Lmarg_a0_phi0} with respect to $t_0$ using \eqref{eq:z} and \eqref{eq:t0_prior}, computing
\begin{multline} \label{eq:Lmarg}
	\int_{a_{0,{\rm min}}}^\infty {\rm d}a_0 P_{\rm prior}(a_0)
		\int_0^{2 \pi} \frac{{\rm d}\phi_0}{2\pi}
		\int_0^T \frac{{\rm d}t_0}{T} \mathcal L(\mathbf{p}, a_0, \phi_0, t_0) \\
	\propto \int_0^T I(\abs{z(\mathbf{p}, t_0)}) {\rm d}t_0
\end{multline}
by numerical quadrature.

Eq.~\eqref{eq:Lmarg} gives the one-detector likelihood marginalized over the nuisance parameters $a_0, \phi_0, t_0$, assuming non-precessing spins and that the dominant mode of GW emission is $(l, m) = (2, \pm 2)$. 
For the case of multiple detectors, the total likelihood is the product of all the one-detector likelihoods, but the marginalization should be made over the source parameters (location, orientation, phase and time of coalescence), that correlate the values of $a_0, \phi_0, t_0$ observed at each detector.

The case of LIGO is particularly extreme because by design, the two detectors at Hanford and Livingston have the same orientation to a good approximation (plus a $90\degree$ rotation in the plane of the detector). Under the approximation that the two detectors are co-aligned, a signal must have the same phase $\phi_0$ and strain amplitude $A_0 \equiv a_0 \sqrt{\langle h_0 \mid h_0 \rangle}$ in both detectors. The arrival times, however, can be different depending on the location of the source, so these must be marginalized over separately. The time delay between detectors is $\delta t = t_d \cos\theta$, where $t_d = \SI{10.012}{\milli\second}$ is the GW travel time between sites and $\theta$ is the angle between the source and a line passing through both detectors. Although an isotropic distribution of sources is uniform in $\cos\theta$ and thus in $\delta t$, the antenna pattern of the detectors induces a selection bias because sources above or below the plane of the detectors are more likely to be detected \citep{Sathyaprakash2009, Chen2017}. The resulting prior for $\delta t$ is well fit by \citep{Cornish2017}
\begin{equation} \label{eq:delta_t_prior}
	P_{\rm prior}(\delta t) \propto 1 - \left( \frac{\delta t}{\SI{10.65}{\milli\second}} \right)^2, 
    	\qquad\abs{\delta t} \leq t_d.
\end{equation}
Since both detectors measure the same polarization, for any given $\delta t$ we can combine the Hanford and Livingston streams of data coherently into a single ``combined channel'', $d_{\rm LIGO}$, in a way that minimises the relative variance:
\begin{equation} \label{eq:d_LIGO}
	\tilde d_{\rm LIGO}(f; \delta t) = \sigma_{\rm LIGO}^2(f) \left(
    	\frac{\tilde d_{\rm H}(f)}{\sigma_{\rm H}^2(f)} 
        - {\rm e}^{-{\rm i} 2\pi f \delta t} \frac{\tilde d_{\rm L}(f)}{\sigma_{\rm L}^2(f)} \right),
\end{equation}
where H and L refer to Hanford and Livingston, and $\sigma_{\rm LIGO}^2 = (\sigma_{\rm H}^{-2} + \sigma_{\rm L}^{-2})^{-1}$ is the noise PSD of the combined channel. The minus sign accounts for the relative $90\degree$ rotation between the two LIGO sites. Since the true value of $\delta t$ is not known, this parameter has to be marginalized over. With the combined channel, we can use the single-detector formulas to get the marginalized likelihood of the LIGO network. From \eqref{eq:d_LIGO} and \eqref{eq:z} we obtain
\begin{equation} \label{eq:z_LIGO}
	z_{\rm LIGO}(\mathbf{p}, t_0, \delta t) = f_{\rm H}^{1/2} z_{\rm H}(\mathbf{p}, t_0) - f_{\rm L}^{1/2} z_{\rm L}(\mathbf{p}, t_0 + \delta t),
\end{equation}
where 
\begin{equation}
	f_\alpha = \frac{\langle h_0 \mid h_0 \rangle_\alpha}{\langle h_0 \mid h_0 \rangle_{\rm H} + \langle h_0 \mid h_0 \rangle_{\rm L}}, \qquad \alpha \in \{{\rm H, L}\}.
\end{equation}
We finally obtain the LIGO-network marginalized likelihood using \eqref{eq:Lmarg}, \eqref{eq:delta_t_prior} and \eqref{eq:z_LIGO}:
\begin{equation} \label{eq:L_LIGO}
	\overline{\mathcal L}_{\rm LIGO}(\mathbf{p})
    \propto \int_0^T {\rm d}t_0 \int_{-t_d}^{t_d} {\rm d}\delta t
    P_{\rm prior}(\delta t) I(\abs{z_{\rm LIGO}(\mathbf{p}, t_0, \delta t)}),
\end{equation}
which we evaluate by quadrature. Note that $a_{\rm 0, min}$ is now interpreted as the detection threshold on the LIGO-network SNR.

Although in this analysis we have been careful to analyse the data of both detectors coherently we have checked that this is largely unimportant for the constraints on the populations we obtain. One gets effectively the same constraints if one treats each detector independently using Eq.~\eqref{eq:Lmarg} and combines them incoherently,
$\log \overline{\mathcal L}_{\rm LIGO}(\mathbf{p}) 
\approx \log \overline{\mathcal L}_{\rm H}(\mathbf{p}) 
+ \log \overline{\mathcal L}_{\rm L}(\mathbf{p})$. The change in the likelihood distribution compared to its width was on the few-percent level or smaller in all cases.

The first five detections reported to date, as well as the last one, are LIGO-only, for which \eqref{eq:L_LIGO} is accurate. The remaining four events were also observed in Virgo. Since Virgo is not co-aligned with LIGO, we cannot apply the treatment above. Instead, we make the approximation that $a_0, \phi_0, t_0$ are uncorrelated between LIGO and Virgo. This amounts to ignore the fact that we know the relative orientations, locations and timing between those detectors. Since we are discarding information, this approximation will increase the uncertainties in $\mathbf p$ without biasing the maximum-likelihood values. For the three-detector events, then, we use
\begin{equation} \label{eq:L}
	\overline{\mathcal L}_{\rm HLV}(\mathbf{p}) \propto \overline{\mathcal L}_{\rm LIGO}(\mathbf{p}) 
        \int_0^T I \left(\abs{z_{\rm V}(\mathbf{p}, t_0)} \right) {\rm d}t_0,
\end{equation}
where V stands for Virgo. As we explained before, treating the detectors independently is sufficient for our purposes even for LIGO, so this is an excellent approximation in the context of our simplified analysis. 

Since we have kept only three parameters, it is practical to evaluate the likelihood \eqref{eq:L} over a grid in $\mathbf{p} = (\mathcal M, q, \chieff)$ for each event. We use a regular grid of $64^3$ points, centered around the values reported by the LVC and with an extent of twice the reported uncertainties (subject to the bounds $0 < q \leq 1, -1 \leq \chieff \leq 1$). In all cases we verify on random parameter values that interpolating the likelihood from the grid has good agreement with the actual computation. By using a grid, we have to compute the likelihood only once, and we are able to apply any prior easily a posteriori. This is a key requirement for model inference, since the prior for the parameter values depends on the population model. Working in a low-dimensional parameter space enables us to circumvent the need for Monte Carlo Markov chains.

\subsection{LIGO--Virgo reported binary black holes}

\begin{figure}
\includegraphics[width=\linewidth]{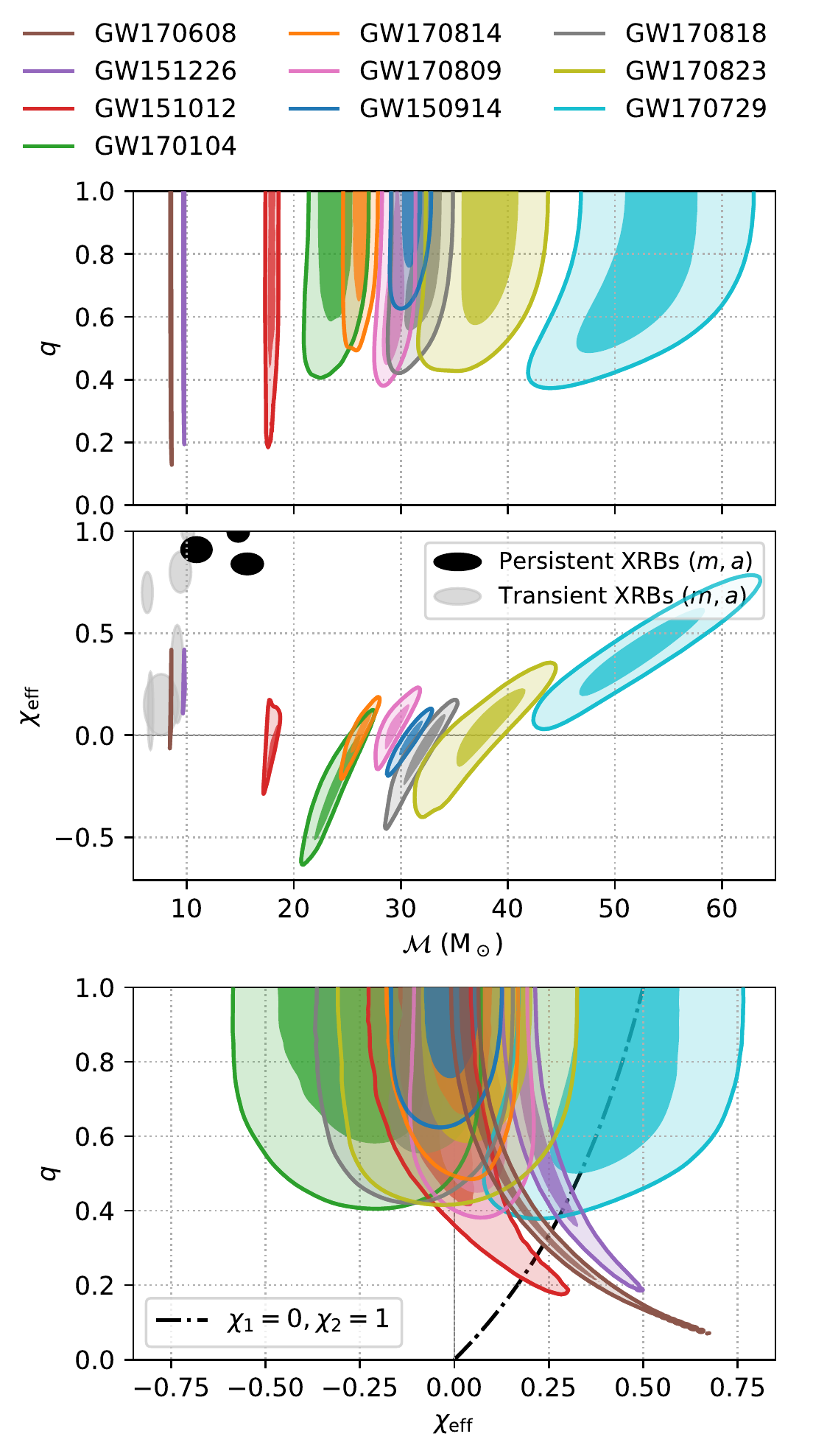}
\caption{Marginalized likelihood contours enclosing $50\%$ and $90\%$ of the distribution for each of the six events reported to date.
In the middle panel, the persistent and transient X-ray binaries reported in \protect\citet{McClintock2013} are shown by ellipses, whose position in the plane represents the black hole mass $m$ and its dimensionless spin $a$.
In the bottom panel, the case where the aligned spins of the black holes are $\chi_1=0, \chi_2=1$ is shown by a dashed-dotted line, as a proxy for what the outcome of a tidally-locked-secondary progenitor would be (see \S\ref{sssec:locked}).}
\label{fig:likelihood}
\end{figure}

The single-event marginalized likelihood computed in this way is shown as a function of the parameters in Fig.~\ref{fig:likelihood}. We obtained very similar results using the \texttt{SEOBNRv4\_ROM} and \texttt{IMRPhenomD} approximants, so we only show the results for \texttt{SEOBNRv4\_ROM}. The likelihood can be interpreted as the posterior distribution that would arise form a uniform prior in $\mathcal M, q, \chieff$. In Appendix~\ref{app:events} we show each event in greater detail, and apply the LIGO prior as a check of our pipeline. 

In the middle panel of Fig.~\ref{fig:likelihood} we superimposed the masses and spins of the black holes in X-ray binaries (XRBs) as reported in \citet{McClintock2013}. The variables for XRBs $(m, a)$ and BBHs $(\mathcal M, \chieff)$ are different, so care has to be taken when comparing them. By definition (Eq.~\eqref{eq:Mchirp_q}), the individual masses of the BBHs are guaranteed to satisfy $m_2 \leq 2^{1/5} \mathcal{M} \leq m_1$, with $2^{1/5}\approx 1.15$. We recall that $\chieff$ is a mass-weighted average of the spin components of the two black holes parallel to the orbital angular momentum. We comment on the XRBs in \S\ref{ssec:spins}.

\begin{figure}
	\centering
	\includegraphics[width=.76\linewidth]{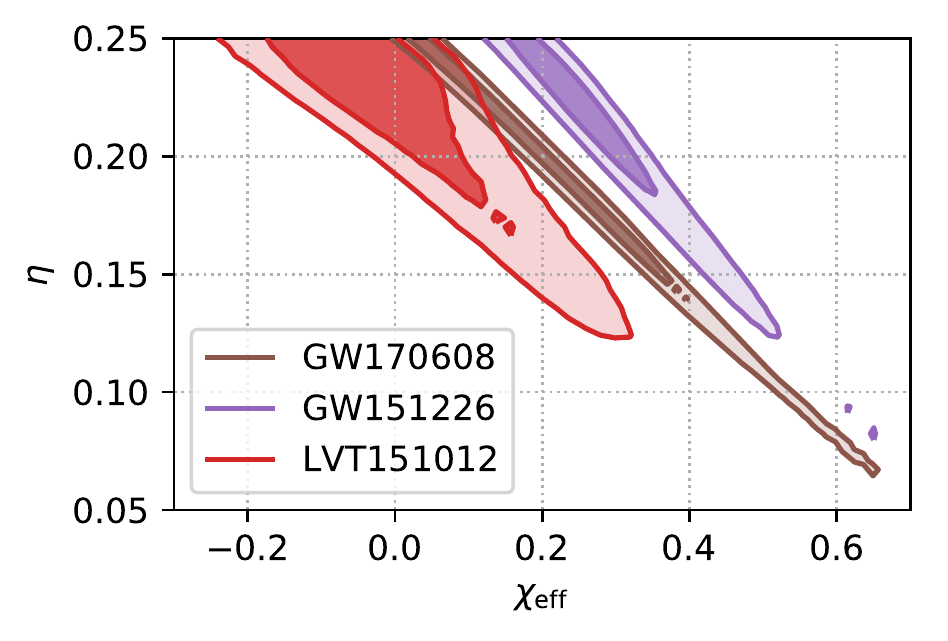}
    \caption{Parameter likelihood for the three lighter likely events, marginalized over $\mathcal M$, as a function of $\eta=q/(1+q)^2$ and $\chieff$. At low mass, $\eta$ and $\chieff$ are degenerate.}
    \label{fig:eta_chieff}
\end{figure}

It is interesting to note that there are clear degeneracies between the parameters and that those degeneracies change with the mass of the system. 
At low mass there is a strong degeneracy between mass ratio and effective spin. This was of course expected (for an early discussion see \citet{PhysRevD.49.2658}). Low mass binaries merge at a higher frequency and thus the detector is more sensitive to the inspiral, where the post-Newtonian (PN) expansion is accurate. The leading PN corrections including spin are approximately degenerate with the leading corrections including mass ratio. This correlation is simpler when expressed using $\chieff$ and the symmetric mass-ratio $\eta = q / (1+q)^2$ as variables \citep{Baird2013, Ng2018}, as we show in Fig.~\ref{fig:eta_chieff} (compare to the bottom panel of Fig.~\ref{fig:likelihood}). It is apparent that a linear combination of the two parameters is better constrained than each of them. We report this combination in Table~\ref{tab:best_constraints}.

At high mass we observe a different degeneracy, between the chirp mass and effective spin (middle panel of Fig.~\ref{fig:likelihood}). 
In General Relativity, the mass can be scaled out of the problem as a time-scale. That is, a system with a smaller chirp mass and all other dimensionless parameters constant yields the same waveform, but sped up, or shifted to higher frequencies. A positive aligned spin can mimic this effect: it acts as an effective repulsive force between the BHs, which delays the plunge and makes them merge at a higher frequency \citep{Campanelli2006}. Since at high masses the chirp mass is not too well constrained, the degeneracy appears. Like before, a linear combination of $\mathcal M, \chieff$ is better constrained than either parameter, which we also report on Table~\ref{tab:best_constraints}.

\begingroup
\renewcommand{\arraystretch}{1.2} 
\begin{table}
  \caption{Best constrained linear combination of $\eta, \chieff$ for the three lighter events, and of $\mathcal M, \chieff$ for the seven heavier ones, with $90\%$ confidence uncertainties. $A$ and $B$ are coefficients chosen to minimise the correlation.}
  \begin{tabular}{l c c c }
    \hline
    Event & $A$ & $\chieff+A(\eta-1/4)$ \\
    \hline
    GW151226 & $2.8$ & $0.16_{-0.03}^{+0.04}$ \\
    GW151012 & $3$ & $-0.09_{-0.1}^{+0.12}$ \\
    GW170608 & $3.4$ & $0.02_{-0.02}^{+0.04}$ \\[1pt]
    \hline
     & $B ~({\rm M_{\sun}})$ & $\mathcal M - B \chieff ~({\rm M_{\sun}})$ \\
    \hline
    GW170104 & $8$ & $25.9_{-1.0}^{+0.8}$ \\
    GW170814 & $8$ & $26.1_{-0.5}^{+0.6}$ \\
    GW170809 & $9$ & $28.9_{-0.6}^{+0.8}$ \\
    GW150914 & $12$ & $31.0_{-0.5}^{+0.6}$ \\
    GW170818 & $11$ & $32.7_{-0.7}^{+1.0}$ \\
    GW170823 & $15$ & $37_{-2}^{+2}$ \\
    GW170729 & $27$ & $41_{-3}^{+3}$ \\[1pt]
    \hline
  \end{tabular}
  \label{tab:best_constraints}
\end{table}
\endgroup

Thus it is more convenient to change parameter basis and use a different combination for the heavy and light events. In the new basis the constraints can be better approximated by a simple Gaussian. Such approximation can provide a quick way to make estimates of the population parameters for astrophysical use.

\section{Model inference}
\label{sec:model_inference}

\subsection{Universe- and detector-rates}
\label{sub:rates}

We turn now to the task of constraining population models combining the data of all events. We will do that by introducing a new set of population-parameters $\bm\mu$, that will depend on the specific model at hand and which we want to constrain. We make two remarks in that respect.

First, we note that what we can constrain is the rate at which BBHs with certain parameter values (masses, spins) merge, which is not necessarily proportional to the abundance of those systems, since their dynamics depend on the parameter values. Light binaries, for example, need to start closer in order to merge in less than the age of the Universe, since GW emission is less efficient than for heavy BBHs. Then, our definition of a model, parametrized by $\bm\mu$, is given by the volumetric merger rate $R(\mathbf{p_s}, z_{\rm rs} \mid \bm{\mu})$, that can depend on the cosmological redshift $z_{\rm rs}$ and the source-frame parameters $\mathbf{p_s} = (\mathcal M_{\rm s}, q, \chieff)$. We recall that $\mathcal M_{\rm s} = \mathcal M/(1+z_{\rm rs})$, while $q, \chieff$ are independent of redshift. $R$ can explicitly depend on redshift if the merger rate depends on time. For example, star formation peaked at $z_{\rm rs} \sim 2$, so if the delays between black hole formation and merger are short compared to the age of the Universe, the merger rate can be expected to be higher at large redshifts. Instead, we will assume the rate to be uniform in comoving volume, so that $R(\mathbf{p_s} \mid \bm{\mu}){\rm d}\mathbf{p_s}$ is the rate at which BBHs with parameter values within ${\rm d}\mathbf{p_s}$ of $\mathbf{p_s}$ merge per unit comoving volume and source-frame time.

Second, the rate of events observed at the detector network is not proportional to the merger rate in the Universe, because of two effects. First, the strength of the GW signal, and thus the horizon distance up to which an event can be observed, again depends on the parameters, inducing a selection bias. And second, the detector- and source-frame masses differ because of cosmological redshift. 
There are additional redshift considerations because GW detectors are sensitive to a luminosity-volume, while we defined the rate per comoving-volume, and because the merger rate is redshifted.
To incorporate these effects, we introduce the detection rate $\lambda$, such that $\lambda(\mathbf{p} \mid \bm{\mu}) {\rm d}\mathbf{p}$ is the rate of detection of events with detector-frame parameters $\mathbf{p}$ per unit time. It is related to the physical rate $R$ by
\begin{multline} \label{eq:lambda}
    \lambda(\mathbf{p} \mid \bm{\mu}) 
    \\
     = 4\pi \int {\rm d}\mathcal M_{\rm s} 
    \int {\rm d}D_c D_c^2 \frac{R(\mathbf{p_s} \mid \bm{\mu})}{1+z_{\rm rs}}
    \delta(\mathcal M - \mathcal M_{\rm s}(1+z_{\rm rs}))
    f(\mathbf{p}, D_L) 
    \\
    = 4\pi \int {\rm d}D_L D_L^2 \frac{1}{(1+z_{\rm rs})^5} 
    R\left(\frac{\mathcal M}{1+z_{\rm rs}}, q, \chieff \,\middle|\, \bm{\mu}\right)
    f(\mathbf{p}, D_L). \\
\end{multline}
Here, $D_c$ is the comoving distance to the source, and $D_L = (1+z_{\rm rs}) D_c$ is the luminosity distance. We have assumed a spatially-flat Universe. Following \citet{Fishbach:2017zga}, we have defined $f(\mathbf{p}, D_L)$ to be the fraction of events with $\mathbf{p}, D_L$ that are detected, averaging over source position $(\theta, \phi)$ and orientation $(\iota, \psi)$.
We take the detection probability to be one if the expectation value of its SNR exceeds a threshold $\rho_{\rm thresh}=9$ and zero if it does not. In practice, the SNR measured by the detector network will differ from its expectation value due to the noise. The measured SNR would be obtained by maximizing $\abs{z_{\rm LIGO}}$ over $t_0, \delta t, \mathbf{p}$; its variance over noise realizations is approximately 1 \citep{Allen2012}. The effect that noise fluctuations have on detectability is important only near the boundary of the sensitive volume, and we ignore it for simplicity. The expectation value of the SNR of an event depends on the angles through
\begin{equation}
	\rho = A(\theta, \phi, \iota, \psi) 
    \frac{\rho_0(\mathbf{p})}{D_{L\rm[Mpc]}},
\end{equation}
where $\rho_0(\mathbf{p}) = \langle h_0(\mathbf{p}) \mid h_0(\mathbf{p}) \rangle^{1/2}$ is the SNR that an optimally aligned source at \SI{1}{\mega\parsec} would have, and the angular factor $0 \leq A \leq 1$ is given e.g. in \citet{Sathyaprakash2009}.
Then,
\begin{equation} \label{eq:f}
  \begin{split}
	f(\mathbf{p}, D_L) &= P(\rho > \rho_{\rm thresh}) \\
    &= P\left(A > \frac{\rho_{\rm thresh}}{\rho_0(\mathbf{p})} 
              D_{L\rm [Mpc]} \right).
  \end{split}
\end{equation}
We estimate $P(A>A_\ast)$ from a histogram of $A$ computed over $10^8$ isotropically distributed realizations of the angles. In Eq.~\eqref{eq:lambda}, we use the redshift-distance relation given in \citet{Adachi2012}, taking the values of the cosmological parameters from \citet{Ade2016}. We evaluate the $D_L$ integral by quadrature.

To compute $\rho_0(\mathbf{p})$, we use a grid over $\mathbf{p}$-space and a reference noise PSD. This is valid provided that the shape of the noise curve of the detectors (i.e. the relative values of the PSD at different frequencies) is approximately constant throughout the observation time, at least in the frequency range relevant for BBH detection.
We construct the reference PSD as the harmonic mean of the combined-channel PSDs of the first six events: $\sigma^2_{\rm ref}(f) = \langle \sigma_{{\rm H},i}^{-2}(f) + \sigma_{{\rm L},i}^{-2} (f) \rangle^{-1}$, where $i$ labels each event and the brackets indicate an average over all the events considered. We consider three BBH events in each observing run, so we expect that this average is representative of the typical PSD during O1 and O2. We find that our results do not sensitively depend on the waveform approximant used. 

\begin{figure}
	\includegraphics[width=\linewidth]{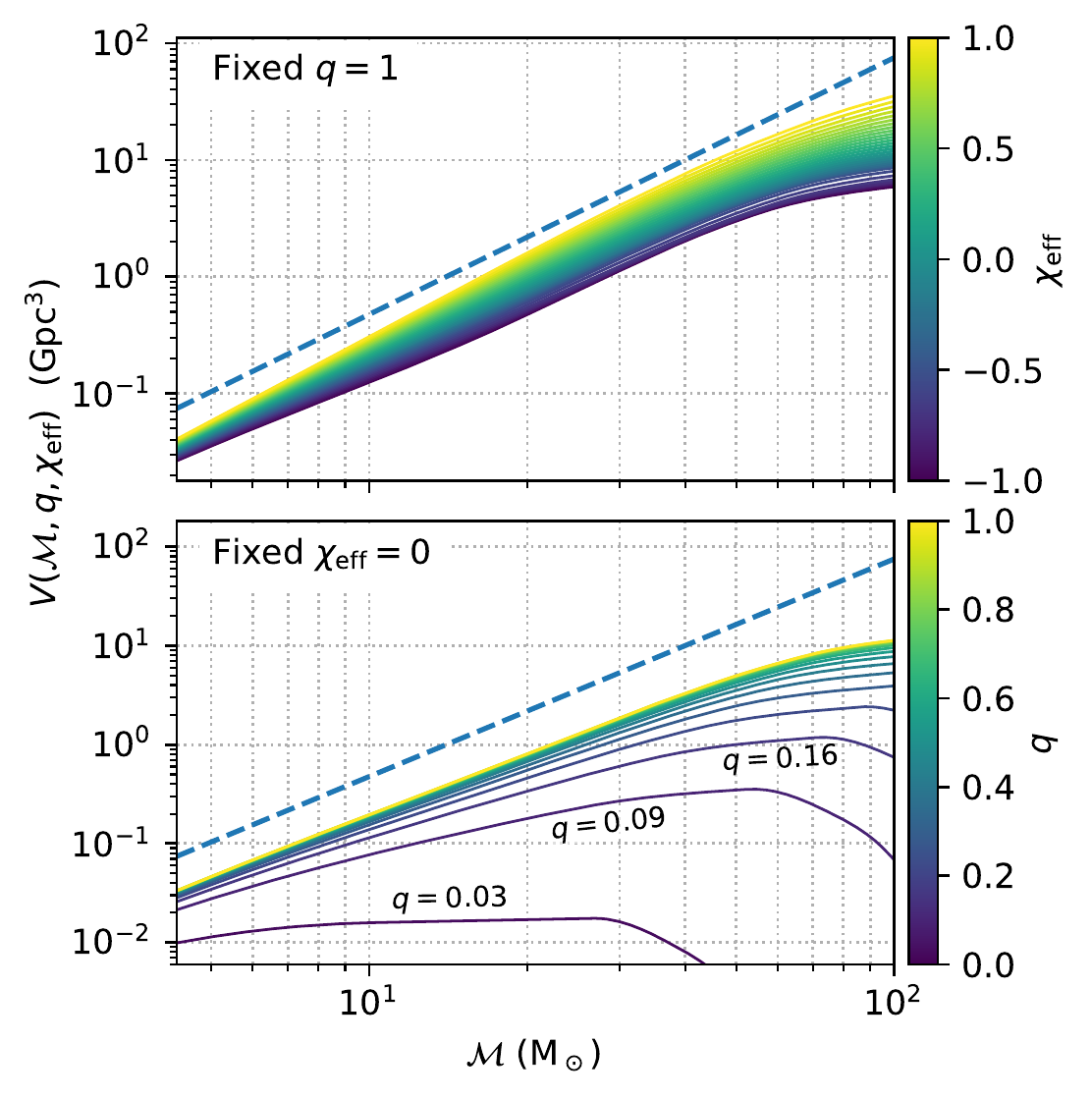}
	\caption{Dependence of the detector network sensitive volume on the parameters $\mathbf{p}=(\mathcal M, q, \chieff)$. The solid lines show $V(\mathcal M)$ for several values of $\chieff$ on a $q=1$ slice (top), or several values of $q$ on a slice of constant $\chieff = 0$ (bottom). The dashed line shows a power law $\propto \mathcal M^{2.2}$ dependence for reference.}
	\label{fig:V}
\end{figure}

If one ignored the fact that the source-frame mass depends on redshift, the detector rate would take the form $\lambda(\mathbf{p} \mid \bm{\mu}) = R(\mathbf{p} \mid \bm{\mu}) V(\mathbf{p})$, where
\begin{equation} \label{eq:V}
	V(\mathbf{p})= 4\pi\int {\rm d}D_L D_L^2 \frac{1}{(1+z_{\rm rs})^4} 
    f(\mathbf{p}, D_L)
\end{equation}
is the sensitive (comoving-) volume of the detector network. 
Fig.~\ref{fig:V} shows the sensitive volume computed with the \texttt{SEOBNRv4\_ROM} approximant. $V$ indeed depends on all the three parameters $\mathcal M, q, \chieff$. As already pointed out in \citet{Fishbach:2017zga}, the mass dependence follows an approximate power law $V \propto \mathcal M^{2.2}$ for $q \gtrsim 0.5$, i.e. heavier BBH mergers are louder. Moreover, events with large $\chieff$ (where the spins are aligned with the orbital angular momentum) are also louder, because $\chieff$ first enters the post-Newtonian expansion as an effective force that is repulsive for $\chieff > 0$ \citep{Campanelli2006}. This effect is irrelevant for the dynamics of the inspiral while the BHs are far apart, but it means that the signal lasts longer in the detector band before the plunge, and thus more SNR is accumulated. Finally, the dependence with $q$ at fixed $\mathcal M$ is very weak if $q \gtrsim 0.5$, but the sensitive volume drops rather strongly for smaller $q$.

Another possible source of bias in the inferred merger rates would arise if the effectualness of the template bank in recovering signals depended sensitively on the parameters. Although this effect is present, its magnitude is much smaller than the sensitive volume dependence for BBHs with $\mathcal M < 100 \,\rm M_{\sun}$ computed here \citep{Canton2017}.

\subsection{Model likelihood}
\label{sub:model_likelihood}

The likelihood that a specific model for the merger rates will have the observed set of detections as an outcome is
\begin{equation} \label{eq:model_likelihood}
  \begin{split}
      P(\{d_i\} \mid \bm{\mu})
          &\propto \prod_{i \in \text{events}}  \int {\rm d}\mathbf{p} P(d_i \mid \mathbf{p}, \mathcal D) P(\mathbf{p}\mid \mathcal D, \bm{\mu}).
  \end{split}
\end{equation}
Here, $P(d_i \mid \mathbf{p}, \mathcal D)$ is the single-event likelihood $\overline{\mathcal L}_i$ given by \eqref{eq:L}, and $P(\mathbf{p} \mid \mathcal D, \bm{\mu})$ is the detection prior for the event parameters $\mathbf{p}$ according to the model $\bm\mu$, which is proportional to the detector rate \eqref{eq:lambda} but normalized to $\int {\rm d}\mathbf{p} P(\mathbf{p} \mid \mathcal D, \bm{\mu}) = 1$.
Using this we can rewrite \eqref{eq:model_likelihood} as
\begin{equation} \label{eq:model_likelihood2}
	P(\{d_i\} \mid \bm{\mu}) \propto 
    	\frac{\prod_i \int {\rm d}\mathbf{p} \overline{\mathcal L}_i(\mathbf{p})
              \lambda(\mathbf{p} \mid \bm{\mu})}
             {\left[ \int {\rm d}\mathbf{p} 
             \lambda(\mathbf{p} \mid \bm{\mu}) \right]^{N_\text{events}}},
\end{equation}
which we can compute on a grid in $\bm \mu$-space, given a merger-rate model $R(\mathbf{p} \mid \bm{\mu})$. The $\mathbf{p}$ integrals in the numerator of \eqref{eq:model_likelihood2} only have support near the measured parameter values of each event, since $\overline{\mathcal L}_i(\mathbf{p})$ is suppressed elsewhere. The integral in the denominator runs over all sensitive parameter space, which we take to be $\mathcal M \in [4.3 \,{\rm M}_{\sun}, 100 \,{\rm M}_{\sun}], q \in [0.03, 1], \chieff \in [-1, 1]$. The lower limit on $\mathcal M$ holds if all astrophysical BHs have $m > 5\,\rm M_{\sun}$. We chose the lower bound on $q$ to include the regions where the reported events have support. Note that such low values of $q$ are outside the calibration region of the approximants \citep{Khan2016,Bohe2017}, although the effect for the purpose of SNR estimation should be minor, and also the approximation that $(l,m)=(2,\pm2)$ is less accurate in that regime. In any case, those mass ratios are highly suppressed in the sensitive-volume weighting (Eq.~\eqref{eq:V}, Fig.~\ref{fig:V}).

To get a posterior distribution for the $\bm\mu$-parameters, one should multiply the likelihood by a prior $P_{\rm prior}(\bm{\mu})$. We will take those priors to be flat. 

We emphasize again that the model likelihood \eqref{eq:model_likelihood2} accounts both for selection effects due to the sensitive volume and for the fact that the priors that one has to use to estimate the parameters of each event depend on the merger-rate model that one is considering.

\section{Astrophysical implications}
\label{sec:results}

With only 10 detections made so far, it is not yet feasible to constrain population models that are too complicated. For this reason, we consider several simple models that intend to probe the different variables separately, and apply Eq.~\eqref{eq:model_likelihood2} to put constraints on their parameters
(see \citet{Taylor2018} for an alternative framework tailored to constraining detailed models).

As before, we repeated all analyses using two different waveform approximants (\texttt{SEOBNRv4\_ROM} or \texttt{IMRPhenomD}). We find that our results are robust to these choices, so below we only show the results for the \texttt{SEOBNRv4\_ROM} approximant.

\subsection{Spin distribution}
\label{ssec:spins}

The distribution of spins of the merger events is currently one of the more informative data the LVC has presented. First, the spin distribution might allow us to distinguish between various formation channels. For example in scenarios where black holes are dynamically captured into binaries one expects each spin to be randomly oriented. For field binaries spins might tend to be aligned with the orbital angular momentum. Tides in binary systems before the second black hole forms might spin up the secondary and align it with the orbital angular momentum. For a chemically homogeneous evolution of the stars to happen, high spins are required. Thus the LVC measurements of $\chieff$ can potentially provide very interesting constraints.

Second, one could try to ascertain whether the properties of the merging black holes are similar to those of black holes in X-ray binaries. In particular there is some indication that local black holes are rotating fast. The middle panel of Fig.~\ref{fig:likelihood} shows the constraints on mass and spin of a collection of black holes in XRBs from X-ray measurements. Heavy black holes in persistent sources, i.e. with heavy companions, which are the natural progenitors of the LIGO/Virgo sources, are close to maximally spinning. Furthermore this spin is usually interpreted as being natal and thus perhaps should apply to the secondary black hole as well. By comparison, the $\chi_{\rm eff}$ reported by the LVC seem rather low. Of course $\chi_{\rm eff}$ constrains only one of the components of the spin and combines both black holes with weights depending on the mass ratio. We will try to use the likelihoods we have computed for the LIGO events to say something about the spin magnitudes and orientations assuming they all come from the same population. 

\subsubsection{Gaussian $\chieff$ rate model}
\label{sssex:gaussian_chi_eff}

In order to understand what the data are already telling us about the distribution of spins we first consider a merger-rate model which is simply a (truncated) Gaussian in $\chieff$, 
\begin{equation} \label{eq:gaussian_chi_eff}
	R(\chieff \mid \overline\chieff, \sigma_{\chieff})
    	\propto G(\chieff - \overline\chieff, \sigma_{\chieff}),
        \qquad \abs{\chieff} < 1;
\end{equation}
we will use $G(x, \sigma)$ to note the Gaussian distribution
\begin{equation*}
	G(x, \sigma) = \frac{1}{\sqrt{2\pi}\, \sigma} \exp(-\frac{x^2}{2 \sigma^2}).
\end{equation*}
We allow a nonzero mean, as expected for example from an isolated-binary formation scenario, and a dispersion $\sigma_{\chieff}$ whose value can help us constrain the typical magnitude of the individual spins. The relevant values of $\sigma_{\chieff}$ turn out to be $\lesssim 0.2$, so in the following we will make no distinction between $\sigma_{\chieff}^2$ and the variance of the truncated Gaussian. For simplicity, in this example we adopt a uniform prior in $\mathcal M, q$. 

\begin{figure}
	\centering
	\includegraphics[width=\linewidth]{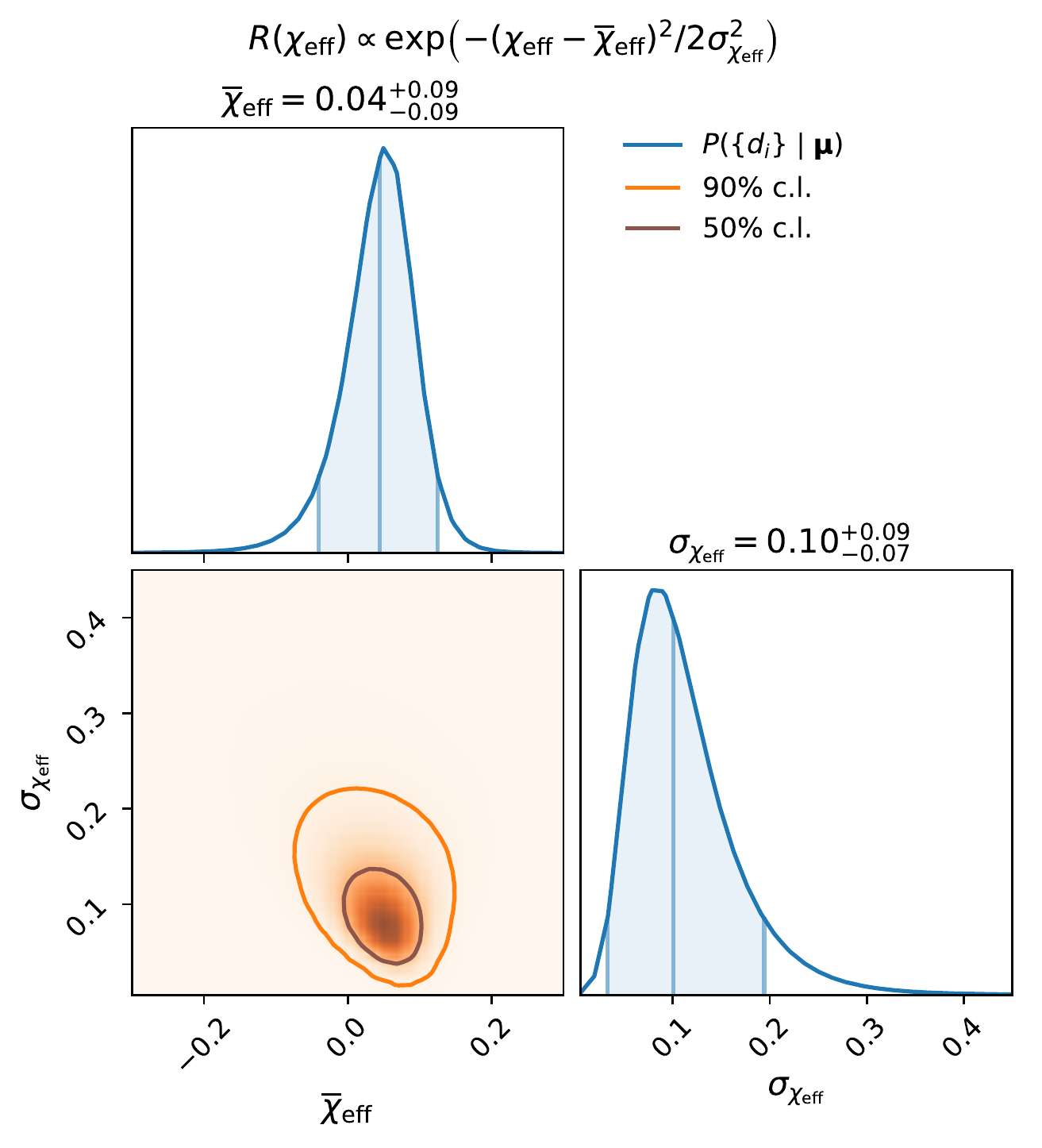}
	\caption{Merger-rate model parameter likelihood for a Gaussian distribution of $\chieff$ given by \eqref{eq:gaussian_chi_eff}. Probability contours enclosing 50\% and 90\% of the distribution are shown in the two-dimensional plot. The one-dimensional plots show the single-parameter marginalized likelihood. The vertical lines show the marginalized distribution median and the minimal 90\% probability interval.}
	\label{fig:gaussian_chi_eff}
\end{figure}

The $\bm\mu$-parameter likelihood is shown in Fig.~\ref{fig:gaussian_chi_eff}. The distribution is consistent with having zero mean, with a mild preference for positive values. The figure also shows that $\sigma_{\chieff}=0$ is inconsistent with the data. We find an upper 90\% bound $\sigma_{\chieff}<0.19$. 

To interpret these results in terms of the distributions of the individual spins we assume that each spin is drawn from a distribution with average spin $\overline a$ and dispersion $\sigma_a$, with an angle relative to the angular momentum whose cosine has a mean $\overline \mu$ and a dispersion $\sigma_\mu$. We will allow the two angles to potentially be correlated so that $\langle \mu_1 \mu_2 \rangle = r_\mu \sigma_\mu^2$. In this case we can compute the mean and variance of $\chieff$:
\begin{equation}
  \begin{split}
    \overline\chieff &= \overline a \,\overline \mu \\
    \sigma_{\chieff}^2 &= \frac{1+q^2}{(1+q)^2} \overline\mu^2 \sigma_a^2 
     + \left( \frac{1+q^2+ 2 q r_\mu}{(1+q)^2} \overline a^2 
     + \frac{1+q^2}{(1+q)^2} \sigma_a^2 \right) \sigma_\mu^2.
  \end{split}
  \label{eq:mean-variance}
\end{equation}

We can first consider situations in which there is no preference for aligned spins, $\overline \mu = \overline{\chieff}=0$ and $\sigma_\mu^2=1/3$, such as BBH populations that would arise in dynamical capture scenarios. The variance of $\chieff$ still depends on $r_\mu$. In the two limits of no correlation ($r_\mu=0$) and perfect correlation ($r_\mu=1$) we get:
\begin{equation}
  \begin{split}
    \sigma_{\chieff}^2 (r_\mu=0) 
         &= \frac{1}{3} \left(\overline a^2 + \sigma_a^2 \right) \frac{1+q^2}{(1+q)^2},\\
    \sigma_{\chieff}^2 (r_\mu=1) 
        &= \frac{1}{3}\left( \overline a^2 + \sigma_a^2 \frac{1+q^2}{(1+q)^2} \right)
  \end{split}
\end{equation}
The dynamical capture scenario would correspond to $r_\mu=0$. 
One can envision a situation for a binary star progenitor in which the spins are misaligned with the orbit but have similar directions, leading to $r_\mu=1$. For example, natal kicks at BH formation if the supernova explosion is asymmetric, or perhaps the tidal interaction with a third body \citep{Rodriguez2018}, would lead to a spin--orbit misalignment.
Spins that are misaligned with the orbit precess, which would spoil the spin--spin alignment. Even so, precession conserves $\chieff$ to a large extent \citep{Apostolatos1994, Racine2008} so it does not affect the $\chieff$ distribution. This holds as long as the orbital angular momentum is bigger than the BH spins, which is generally the case unless the mass ratio is very small, $q \lesssim \chi_1 v/c$ (so it is valid for the sources LIGO/Virgo are most sensitive to, see Fig.~\ref{fig:V}).
The limit of completely random misalignment between spin and orbital angular momentum considered here is extreme, for example the necessary kicks would unbind most systems. In any event we consider it to understand the limits of the constraints we get. 

We first consider whether the data allow for high spins as might be hinted by XRBs. For equal mass ratios and assuming $\overline a \approx 1$, these dispersions would be $\sigma_{\chieff}(r_\mu = 0, q = 1)= 1/\sqrt 6 \approx 0.4$ and $\sigma_{\chieff}(r_\mu=1, q=1)= 1/\sqrt 3 \approx 0.6$. Both values appear to be too large, indicating that the data already do not favour large spins even if both of them are randomly oriented. One could consider small mass-ratios, in which case for high spins the dispersions become $\sigma_{\chieff}(r_\mu = 0, q \ll 1)\sim \sigma_{\chieff}(r_\mu = 1, q \ll 1) \sim 1/\sqrt 3 \approx 0.6$, which again seems disfavoured. 

In the binary progenitor scenario it is believed that the BHs have a preference for being aligned with the orbital angular momentum. As an extreme version let us consider the case of perfect spin--orbit alignment $\overline\mu = 1, \sigma_\mu = 0$. The mean and variance become:
\begin{equation}
  \begin{split}
	\overline\chieff &= \overline a \\
    \sigma_{\chieff}^2 &= \frac{1+q^2}{(1+q)^2} \sigma_a^2.
  \end{split}
\end{equation}
Fig.~\ref{fig:gaussian_chi_eff} shows that the data are consistent with having zero mean for $\chieff$, and at most around $\overline{\chieff}\lesssim 0.1$, thus $\overline a \lesssim 0.1$. The data also require a non-zero variance, this demands $\sigma_a \neq 0$ (because we have assumed perfect alignment, $\sigma_\mu=0$ and the only source of variance for $\chieff$ in Eq.~\eqref{eq:mean-variance} remains $\sigma_a$). Since $a>0$, in this limit of small spins one cannot have $\sigma_a \gg a$. Thus, from Fig.~\ref{fig:gaussian_chi_eff} the only viable region for the aligned scenario is $\sigma_a \sim \overline a \sim 0.1$

\subsubsection{Rate model for individual spins}

Although we have already obtained most of the interesting physical conclusions from the previous simple analysis perhaps it is warranted to be a bit more careful with the dependence on $q$. Both the model predictions Eq.~\eqref{eq:mean-variance} and the inferred $\chi_{\rm eff}$ for the events (Fig.~\ref{fig:likelihood}) depend on $q$. In the case of the data, for the lower-mass events there is a very strong degeneracy between $\chieff$ and $q$. It is only for small mass-ratios that the spin parameter becomes large. Therefore, when we look at the constraints on $\sigma_{\chi_{\rm eff}}$ larger values are allowed due to the possibility that the events have small $q$. But the models tend to predict a larger variance in this regime, because in this limit only one of the two spins contributes to $\chieff$ and there is no possibility of cancellation. Thus, it is worth considering directly a prior on the individual spins rather than on $\chieff$ so that the $q$ dependence is automatically incorporated. 

With this small number of events we do not want to consider very complicated rate models. We restrict ourselves to a two-parameter model that explores at the same time the typical value of the individual spin magnitudes as well as their alignment with the orbit. We will consider a rate model where each spin--orbit alignment $\mu$ is uniformly distributed between $\mu_{\rm min}$ and 1. For the spin magnitude we know that in the limit of perfect alignment the data require that the spin distribution have some variance. Thus we will consider a uniform distribution between $a_\ast - \Delta_a$ and $a_\ast + \Delta_a$ and move $a_\ast$. When $a_\ast$ approaches the boundaries only $0<a<1$ is allowed. We will take $\Delta_a=0.1$. 

\begin{figure}
\includegraphics[width=\linewidth]{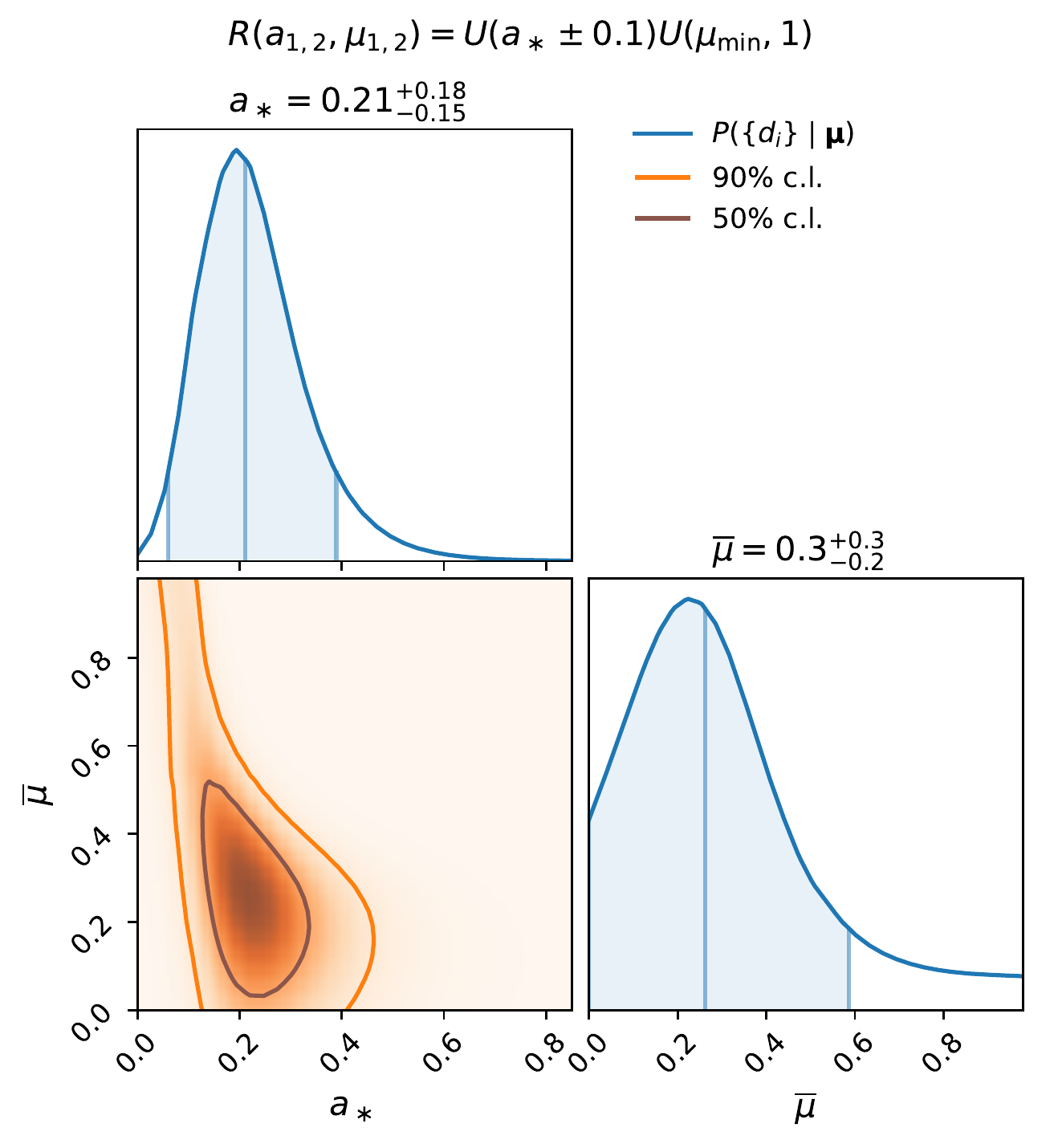}
\caption{Parameter likelihood for a merger-rate model in which the BBHs have spin magnitude $a$ uniformly distributed between $a_\ast-0.1$ and $a_\ast+0.1$, and alignment between the BH spin and the orbital angular momentum uniformly distributed between $\mu_{\rm min}=1-2 \overline \mu$ and $1$.}
\label{fig:a_bar_mu_avg}
\end{figure}

The likelihood for $a_\ast, \overline\mu$ is shown in Fig.~\ref{fig:a_bar_mu_avg}. We find that the typical spin magnitudes have a preferred value $a_\ast \approx 0.2 \pm 0.2$, in agreement with \citet{Wysocki2018}. The alignment has an upper $90\%$ bound of $\overline\mu \lesssim 0.6$, but there is support all the way up to $\overline\mu = 1$ if the typical spins are $a_\ast \approx 0.1$, consistent with our estimate in \S\ref{sssex:gaussian_chi_eff}.

\subsubsection{Tidally-locked progenitor} \label{sssec:locked}

We now consider the effects of tides in the binary. For a field binary, before becoming a BBH the progenitor was a binary star where the companion of the primary BH probably was a Wolf--Rayet star  (the core of a star that lost its envelope). The minimum initial distance required for the BBH to merge in the age of the Universe due to the emission of gravitational radiation is comparable to the minimum distance required to tidally lock the companion star within its lifetime, and the time-scales for these two processes are very steep functions of the distance, so two distinct subpopulations are expected \citep{Kushnir:2016zee,Hotokezaka:2017esv,Zaldarriaga:2017qkw,Qin2018}. In systems where the merger time is shorter than $\sim 10^8$ years, the secondary would be tidally locked and rapidly spinning. If the merger time is longer, tides are too weak. We might thus expect two distinct populations. 

A strong natal kick to the second BH could misalign the orbit, but we ignore this case as it is rather unlikely given that at the time of the second explosion the binary is already tight, and thus the needed kick velocity to produce large misalignments is too large. We assign the high spin to the secondary (lighter) BH. 

\begin{figure}
	\centering
    \includegraphics[width=\linewidth]{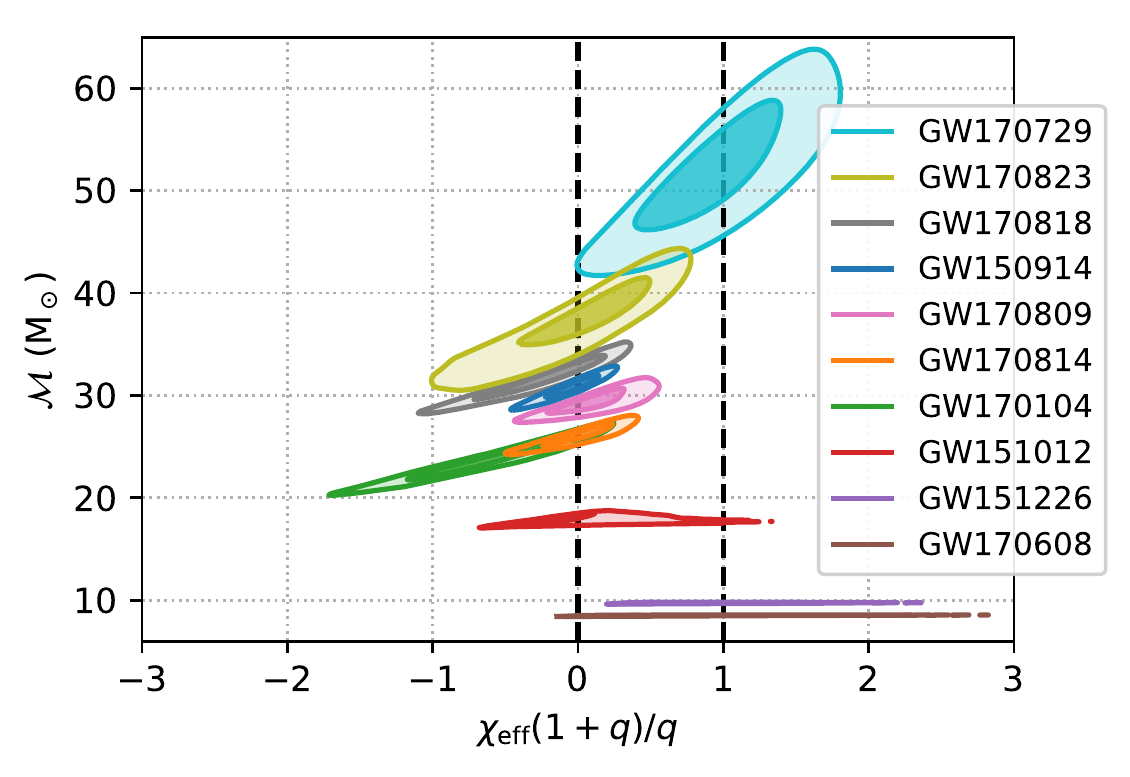}
    \caption{Single-event likelihood in terms of $\chieff$ rescaled by the tidally-locked-progenitor proxy $q/(1+q)$, and marginalized over $q$. This channel predicts a bimodal distribution in this variable, with peaks at 0 and 1.}
    \label{fig:renormalized_chi}
\end{figure}

In Figs.~\ref{fig:likelihood} and \ref{fig:events}, \ref{fig:GW170729} we used the prescription $\chi_1=0, \chi_2=1$ as a proxy for the case in which the Wolf--Rayet star is successfully locked, which defines the curve $\chieff = q/(1+q)$ shown. We find that GW170729, GW151226, GW151012 and GW170608 are consistent with this proxy. The latter three are consistent if their mass ratio is low, while all the other, heavier events are more consistent with $q\sim 1$. Thus this would be slightly unlikely if all the events belong to the same population. With the current error bars and small number of events this is at most a qualitative hint. To further illustrate this, in Fig.~\ref{fig:renormalized_chi} we show the likelihood as a function of a rescaled $\chieff$ in units of the proxy $q/(1+q)$. The rescaled effective spin can take values between $\pm(1+q)/q$, which for a reference $q=1/2$ is $\pm 3$. \citet{Zaldarriaga:2017qkw} predicted a bimodal distribution in this variable, with peaks at 0 and 1 corresponding to the two subpopulations.

We implement a two-parameter model of this scenario, described by the fraction $f$ of systems where the progenitor has been tidally-locked, and the standard deviation $\sigma_{\chieff}$ in the effective spin of the other subpopulation. 
We assume that the aligned spin of each black hole either comes from the same zero-mean Gaussian distribution, or is unity for a fraction $f$ of the secondaries. That is, the standard deviation of the distributions of individual spin and effective spin are related by
\begin{equation}
	\sigma_{\chi_1} = \sigma_{\chi_2} = \frac{1+q}{\sqrt{1+q^2}}\sigma_{\chieff},
\end{equation}
valid for the black holes that do not come from a tidally-locked progenitor.
This results in a superposition of two Gaussian distributions for $\chieff$:
\begin{multline} \label{eq:R_locked}
	R(\chieff, q \mid f, \sigma_{\chieff})
    	= (1-f)G(\chieff, \sigma_{\chieff})\\
          + f G \bigg(\chieff - \frac{q}{1+q}, 
                      \frac{\sigma_{\chieff}}{\sqrt{1+q^2}} \bigg).
\end{multline}
The second term describes the locked-progenitor subpopulation, it has a positive mean due to the maximally-spinning secondary, and a smaller variance because in this case only the primary is random.

It should be noted that varying the prescriptions used for stellar winds and tides it can also be possible to obtain intermediate distributions where the secondary may not be maximally spinning after being tidally locked \citep{Qin2018}. However, it would be hard to constrain models more complicated than \eqref{eq:R_locked} with the present number of events; and the upper bound on $f$ would get stronger as long as we interpret $f$ as the fraction of maximally-spinning secondaries.

\begin{figure}
\centering
\includegraphics[width=\linewidth]{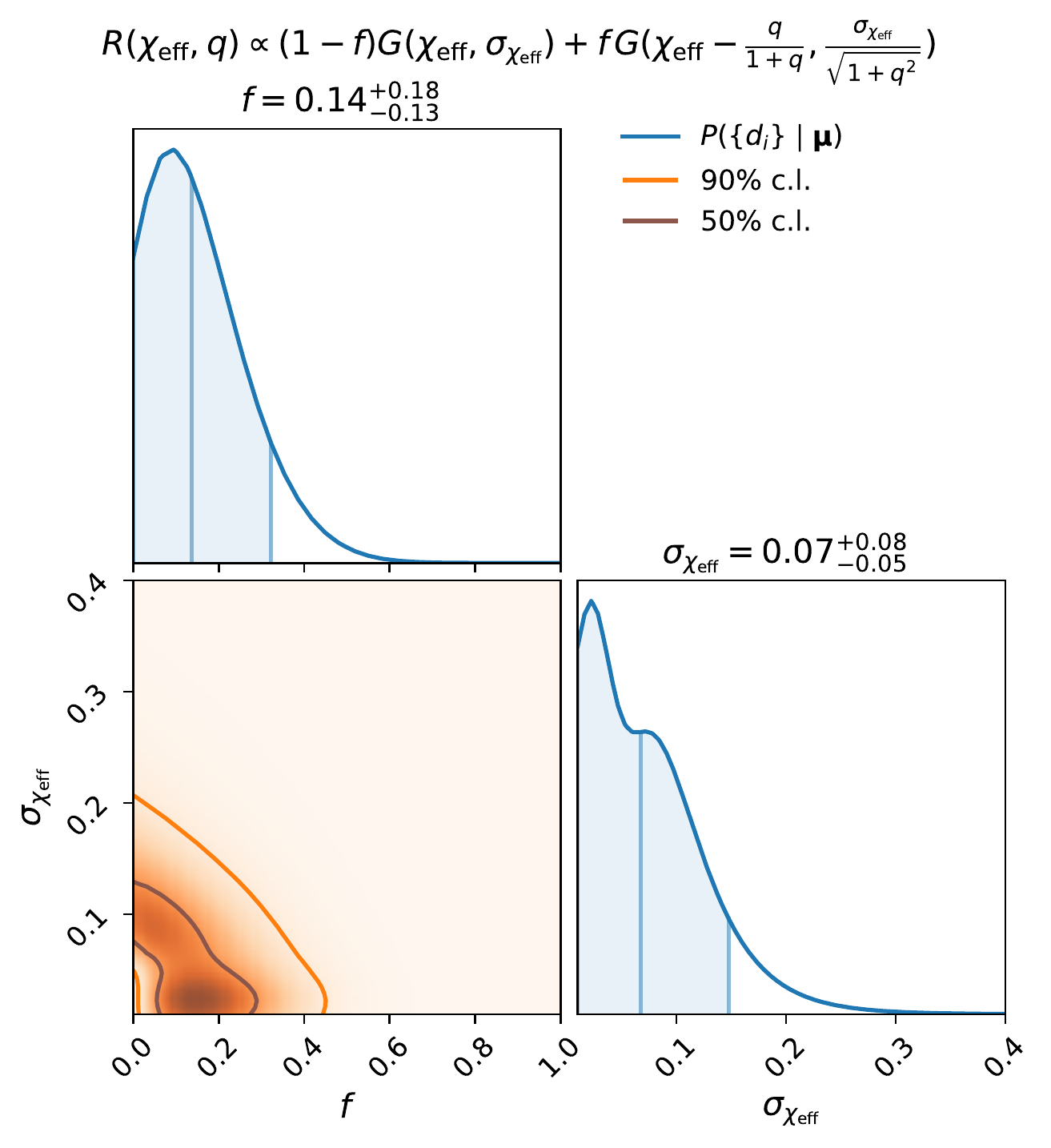}
\caption{Parameter likelihood for a merger-rate model where a fraction $f$ of the sources comes from a tidally-locked progenitor. The distribution of $\chieff$ is given by \eqref{eq:R_locked}.}
\label{fig:gaussian_2ndlocked_chi_eff}
\end{figure}

The likelihood for $f, \sigma_{\chieff}$ is shown in Fig.~\ref{fig:gaussian_2ndlocked_chi_eff}. We see that the data are consistent with $f=0$ and puts an upper 90\% bound $f < 0.3$. One can see a hint of two peaks, one in which the random spin is used to explain all the events, leading to a larger variance, and the other where the higher-spin events are explained using tides and the random component of the spins is low. The current number of events is too small to discriminate but it should become possible in the next LIGO run. Furthermore if a negative $\chieff$ were to be observed, especially a large one compared with the width of the random component, then this scenario would be disfavoured.

\subsubsection{Trends with mass and GW170729} \label{sssec:trend-mass}

It is important to notice that in the analysis so far we have assumed that all the LIGO events are samples from a unique population. Fig.~\ref{fig:likelihood} shows some noticeable differences between the light and heavy systems. The most obvious difference is just related to the parameter combination than can be measured best, the change in the degeneracy direction from being between spin and mass ratio to being between spin and mass. Of course this is just a feature of the way the strain depends on the parameters. 

However, prior to the detection of GW170729, a hint that the lightest systems might be the ones with a clearer positive spin was apparent. With such small number of events we did not dare start splitting the sample in subsets and especially in such an a posteriori fashion. However, a trend with mass could be expected on theoretical grounds and is sometimes used as input in population synthesis models (see for example figure 1 of \citet{Belczynski:2017gds}).
Of course one has to worry about those inputs because even though they might be physically motivated, some of them were chosen after seeing the first set of LIGO BBHs.

GW170729 is heavy and has a moderately high spin, which breaks this potential trend and could mean either that there is no such trend or that a different physical mechanism originated its spin. Formation scenarios that would naturally account for this event include tidal-locking after a common-envelope phase as discussed in \S\ref{sssec:locked}; a chemically-homogeneous formation, which predicts masses in the range of GW170729, near-equal mass ratio, aligned spins and a peak merger rate at redshift 0.5 \citep{Mandel2016}; or a repeated merger within the globular cluster scenario, which would also have high mass and spin (although randomly aligned) and would represent a fraction of the mergers coming from globular clusters that could be as high as 20\% if the spins at birth are small \citep{Rodriguez2018b}.

\subsection{Mass distribution}

Another interesting question is the distribution in mass and the potential lack of heavy systems. This was already pointed out in \citet{Fishbach:2017zga}. Here we repeat that analysis with six additional events, allowing for spin (through $\chieff$ only) and including the source-frame mass dependence on redshift. For simplicity we directly model the distribution as a function of $\mathcal M$ and adopt a power-law prior with a cutoff. One could model the distribution of the individual masses and make further assumptions about how correlated the two masses are. We feel that this is unnecessary at this stage as the events have mass ratios $q\sim 1$ and a cutoff in the mass distribution would also lead to a cutoff in the chirp mass distribution at a very similar mass, especially given the current errors and small number of events. We adopt a $5\,{\rm M_{\sun}}$ low-mass cutoff and a free high-mass cutoff parameter $\mathcal M_{\rm max}$ for the model distribution. We recall that we assumed that the detector network is sensitive to events with $\mathcal M < 100\,\rm M_{\sun}$, which is similar to and less conservative than the $m_1+m_2 < 200 \,\rm M_{\sun}$ case analysed in \citet{Fishbach:2017zga}.

\begin{figure}
\centering
\includegraphics[width=\linewidth]{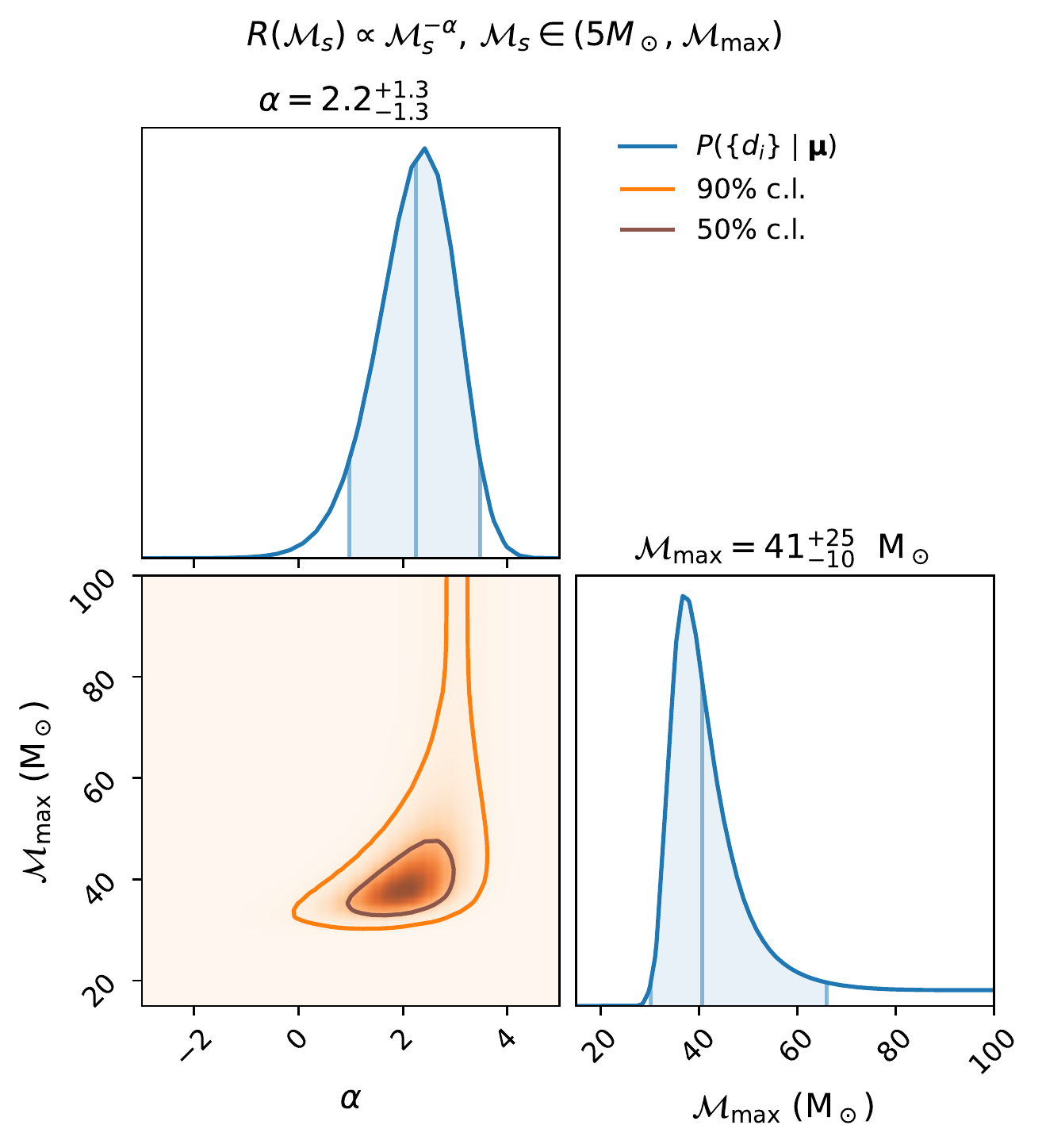}
\caption{Parameter likelihood for a merger-rate model where the source-frame chirp-mass distribution follows a power law $\mathcal M_s^{-\alpha}$, with a low-mass cutoff at $5\,{\rm M_{\sun}}$ and a high-mass cutoff at $\mathcal M_{\rm max}$.}
\label{fig:M_chirp_power_law_cutoff}
\end{figure}

We show our results in Fig.~\ref{fig:M_chirp_power_law_cutoff}, which qualitatively agree with \citet{Fishbach:2017zga} and give a preliminary indication of the presence of a maximum mass. The distribution median is $\mathcal M_{\rm max}=41 \,\rm M_{\sun}$, which would correspond to $m_{1,2} = 47 \,{\rm M_{\sun}}$ if $q=1$, also in agreement with \citet{O2populations}. The lower bound on $\mathcal M_{\rm max}$ is determined by the heaviest event detected, GW170729.

One potential caveat is that the glitch background in LIGO might increase with mass, partially compensating the increase in volume up to which heavy systems can be seen. To investigate this we estimate how sensitive our results are to an increase in the background. From fig.~9 of \citet{O1}, the background of triggers during O1 is a steeply decreasing function of the detector network SNR $\rho$ and it is approximated by a straight line for $\log_{10} N_{\rm bg}(\rho)$. This also holds for unmodeled searches \citep{Lynch2018}. The background is dominated by detector glitches. GW signals from heavy BBHs have short duration and fewer cycles in the detector band, so glitches can more easily resemble them. The background level is then dependent on the mass scale and a stricter threshold on the SNR might be necessary for heavier events, which has the effect of reducing the sensitive volume relative to our previous estimate \eqref{eq:V}. We can make a simple estimate of this effect as follows. In an Euclidean spacetime (setting $z_{\rm rs}(D_L)=0$ in Eq.~\eqref{eq:V}), the sensitive volume scales as $V(\mathbf{p}) \propto \rho_0^3(\mathbf{p})$, since the SNR of a source decays as $D^{-1}$. If we allow for a parameter-dependent threshold on the SNR, $V$ gets a correction
\begin{equation}\label{eq:V_corrected}
	\frac{V(\mathbf{p})}{V(\mathbf{p_0})} = 
    	\left( \frac{\rho_0(\mathbf{p})}{\rho_0(\mathbf{p_0})} \right)^3
    	\left( \frac{\rho_{\rm thresh}(\mathbf{p_0})}{\rho_{\rm thresh}(\mathbf{p})} \right)^3,
\end{equation}
where $\mathbf{p_0}$ is some reference parameter value. The first term in the right hand side comes from Eq.~\eqref{eq:V}, and the second describes the reduction in sensitive volume if the threshold $\rho_{\rm thresh}(\mathbf{p}) > \rho_{\rm thresh}(\mathbf{p_0})$. In an expanding Universe, the $(1+z_{\rm rs})^{-4}$ term in Eq.~\eqref{eq:V} suppresses the large-luminosity-distances contribution to $V$, so the relative decrease in sensitive volume due to raising the threshold for large masses is smaller than the estimate \eqref{eq:V_corrected}.

As an example, if the glitch rate increased from the one reported in fig.~9 of \citet{O1} by a factor of $100$ above a certain chirp mass, the SNR threshold would have to be raised by $\Delta \rho_{\rm thresh} \approx 1.07$ for those events to have the same false-alarm rate. Assuming a detection threshold of $\rho_{\rm thresh}(\mathbf{p_0}) = 9$, the sensitive-volume correction factor in Eq.~\eqref{eq:V_corrected} is $0.71$. There is no indication of such a dramatic increase in the background even when going beyond $\mathcal M > 100 \,\rm M_{\sun}$ \citep{Abbott:2017iws}. A thorough test with an injection campaign was performed by \citet{O2populations} and found correction factors consistent with this estimate. Thus we believe the hint of a cutoff mass to be a robust result.

\subsection{Mass-ratio distribution}

\begin{figure}
\centering
\includegraphics[width=.61\linewidth]{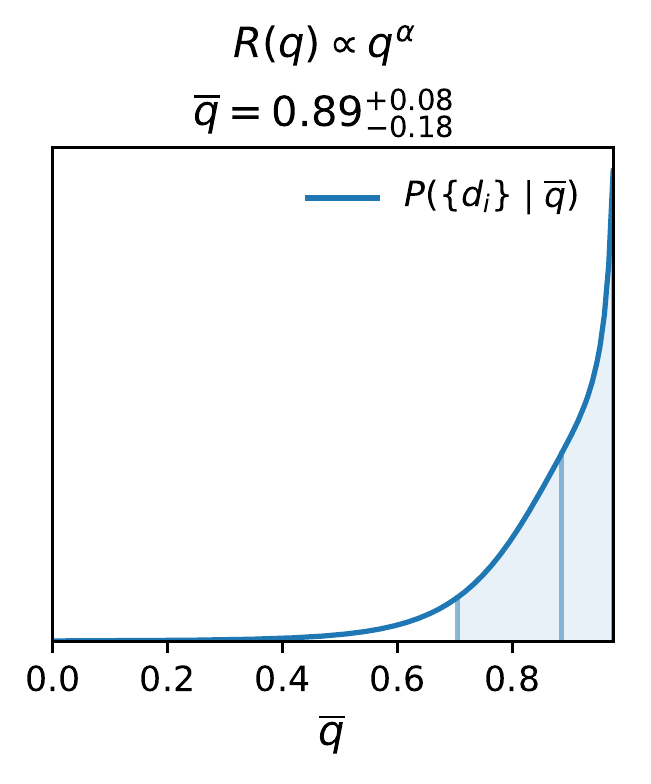}
\caption{Parameter likelihood for the average mass-ratio $\overline q$, in a merger-rate model where the mass-ratio distribution follows a power law $q^\alpha$, with $\alpha$ adjusted to yield $\langle q \rangle = \overline q$.}
\label{fig:q_power_law}
\end{figure}

Finally, we consider what we can say about the mass-ratio distribution. The likelihoods themselves are rather flat in $q$ so we do not expect particularly good constraints. We consider a power law in $q$ and try to constrain the exponent. We present our results in Fig.~\ref{fig:q_power_law}. We find that typical mass ratios below $0.7$ seem disfavoured. For comparison, the average mass ratio of a distribution where the binary masses are independently taken from a power-law $P(m) \propto m^{-\alpha}$ is $\overline q = 1-1/\alpha$. From Fig.~\ref{fig:M_chirp_power_law_cutoff} we obtain that $\alpha\approx 3$ is favoured by the data if we impose no cutoff ($\mathcal M_{\rm max} \to \infty$); independent draws from this distribution which would yield $\overline q = 2/3$, consistent with our lower bound on $\overline q$. If we use the Salpeter mass function $\alpha = 2.35$, then $\overline q = 0.575$, which is disfavoured. A uniform distribution in $q$ would have $\overline q=1/2$, also disfavoured.
We note that the statement that equal mass-ratios are favoured holds for the physical merger-rate in the Universe, even after accounting for selection effects due to the higher sensitive volume of the detector to those systems.

Due to the degeneracy between spin and mass ratio we could consider the combination of a Gaussian in $\chieff$ and a power law in $q$. However, when this is done we find the constraints on the variance of the Gaussian and the mean $\overline q$ to be the same as those we report in Figs.~\ref{fig:gaussian_chi_eff} and \ref{fig:q_power_law}.

\section{Discussion}

It is clear that even with the small number of events reported so far the LIGO--Virgo data already provide hints that have interesting astrophysical consequences. Most interestingly, the data suggest that the BHs do not all spin rapidly, perhaps in tension with the simplest interpretation of the X-ray binary results. This is even true in dynamical scenarios where the spins are randomly oriented and even more so in the field binary case where there might be a tendency towards alignment between spins and angular momentum.

These results are largely consistent with those presented in \citet{Farr2017}, where aligned high spin distributions were disfavoured. Quantitatively we differ because we have more events, have a better estimate of the likelihoods for each event including the degeneracy between mass ratio and spin, and rather than comparing discrete models we have continuous parameters that connect them, allowing us to perhaps have a better sense of the typical spins that are favoured or disfavoured in individual scenarios.

There are many potential explanations for the difference between the typical spins of high-mass XRBs and the LIGO/Virgo systems. Of course both data sets contain a small number of events and thus this could be a statistical fluke. Perhaps there is an unrecognized systematic in one or both measurements. There could be astrophysical explanations, perhaps these systems come from different populations. There could even be exotic explanations such as the effect of an axion-type particle that through superradiance extracts energy from rotating black holes to produce a cloud of axions around them. 

At the current time the simple test of seeing if the spin distribution has a tendency towards positive spins is not powerful enough. We do not detect a mean to the distribution and thus cannot use that to distinguish between field binary and dynamical scenarios. Regarding tides, four of the events are consistent with having a maximally spinning secondary, although for the lighter three this only happens if the mass ratio for these events is lower than for the heavy ones, and in particular GW151012 is only marginally consistent. With the data at hand we cannot make a conclusive statement but again it is clear that the next few events are going to be very interesting. 

Regarding the mass distribution we confirm previous findings that there is a hint of a cutoff in mass. The distribution of mass ratios is not so well constrained, in part due to the degeneracy with the spin, but distributions where the BBHs have comparable masses are favoured.

The code and data used in this work are publicly available at \url{https://github.com/jroulet/constraints\_bbh\_populations}.

\section*{Acknowledgements}
We thank to Ben Bar-Or, Liang Dai, Vera Gluscevic, Tejaswi Venumadhav and Barak Zackay for useful discussion, and Christopher Berry, Riccardo Sturani and Salvatore Vitale for comments on the manuscript. We are grateful to the LVC for releasing the strain data and software through the GWOSC. M.Z. is supported by NSF grants AST-1409709 and PHY-1521097, the Canadian Institute for Advanced Research (CIFAR) program on Gravity and the Extreme Universe and the Simons Foundation Modern Inflationary Cosmology initiative.
This research has made use of data and software obtained from the Gravitational Wave Open Science Center (\url{https://www.gw-openscience.org}), a service of LIGO Laboratory, the LIGO Scientific Collaboration and the Virgo Collaboration. LIGO is funded by the U.S. National Science Foundation. Virgo is funded by the French Centre National de Recherche Scientifique (CNRS), the Italian Istituto Nazionale della Fisica Nucleare (INFN) and the Dutch Nikhef, with contributions by Polish and Hungarian institutes.

\bibliographystyle{mnras}
\bibliography{references}

\appendix
\section{Selection bias} \label{app:bias}

In this appendix we discuss in further detail the equivalence between our derivation of Eq.~\eqref{eq:model_likelihood2} and others present in the literature. Letting $\mathcal D$ be the proposition that the event was detected (and $\neg \mathcal D$ its negation), we can write the marginalized likelihood as
\begin{equation} \label{eq:P(d|p)app}
  \begin{split}
      P(d \mid \mathbf{p})
      &= P(d \mid \mathcal D, \mathbf{p}) P(\mathcal D \mid \mathbf{p})
        + P(d \mid \neg \mathcal D, \mathbf{p}) P(\neg \mathcal D \mid \mathbf{p}) \\
      &= P(d \mid \mathcal D, \mathbf{p}) P(\mathcal D \mid \mathbf{p})
        + P(\neg \mathcal D \mid d, \mathbf{p}) P(d \mid \mathbf{p}) \\
      &= P(d \mid \mathcal D, \mathbf{p}) P(\mathcal D \mid \mathbf{p}).
  \end{split}
\end{equation}
The last equality follows because the data segments we analysed were detections, and the criterion for detection depends only on the data, so $P(\neg \mathcal D \mid d) = 0$. The term $P(\mathcal D \mid \mathbf{p})$ is the observational bias and can be computed as
\begin{equation} \label{eq:P(D|p)}
  \begin{split}
    P(\mathcal D \mid \mathbf{p})
      &= \int_{a_{0, \rm min}}^\infty P(a_0 \mid \mathbf{p}) {\rm d}a_0 \\
    &= \int_0^{D_h(\mathbf{p})} P(D) {\rm d}D \\
    &\propto V(\mathbf{p}),
  \end{split}
\end{equation}
where $D_h(\mathbf{p})=\sqrt{\langle h_0 \mid h_0 \rangle (\mathbf{p})}\,\SI{}{\mega\parsec}/a_{0,\rm min}$. The probability of detection depends on the parameters because we are keeping only a subset $\mathbf{p}$ of the parameters, and in particular marginalizing over distance and angles (cf. \citet{Loredo2004,Mandel2018}, keeping all the parameters would lead to a detection probability of 1 so they can omit this term).
Using Eqs.~\eqref{eq:P(d|p)app} and \eqref{eq:P(D|p)} we can readily show the equivalence between our Eq.~\eqref{eq:model_likelihood2} and eq.~(7) of \citet{Mandel2018}. The expressions differ only in the integrands of the numerators of each term in the product:
\begin{equation} \label{eq:pop}
  \begin{split}
    P(d_i \mid \mathbf{p}, \mathcal D_i)
      \lambda(\mathbf{p} \mid \bm{\mu})
    &= P(d_i \mid \mathbf{p}, \mathcal D_i)
      V(\mathbf{p})R(\mathbf{p} \mid \bm{\mu}) \\
    &\propto P(d_i \mid \mathbf{p}, \mathcal D_i)
      P(\mathcal D \mid \mathbf{p}) R(\mathbf{p} \mid \bm{\mu}) \\
    &= P(d_i \mid \mathbf{p}) R(\mathbf{p} \mid \bm{\mu})
  \end{split}
\end{equation}
which is the form found in \citet{Mandel2018}.
Note that Eq.~\eqref{eq:pop} is the posterior for $\mathbf{p}$ under a prior labeled $\bm\mu$. In Fig.~\ref{fig:events} we use the LIGO prior as an example.

We have expressed Eq.~\eqref{eq:model_likelihood2} in terms only of $P(d \mid \mathbf{p}, \mathcal D)$ and $\lambda(\mathbf{p} \mid \bm{\mu})$ instead of the more physically-interesting quantities $P(d \mid \mathbf{p})$ and $R(\mathbf{p} \mid \bm{\mu})$. Our motivation is that this expression is perhaps more natural from the perspective of the observers, since the outcome of the observations can only depend on the event rate at the detector and the events analyzed will necessarily be conditioned to detection.

\section{Comparison with LVC results}\label{app:events}

\begin{figure*}
\centering
\includegraphics[width=.331\linewidth]{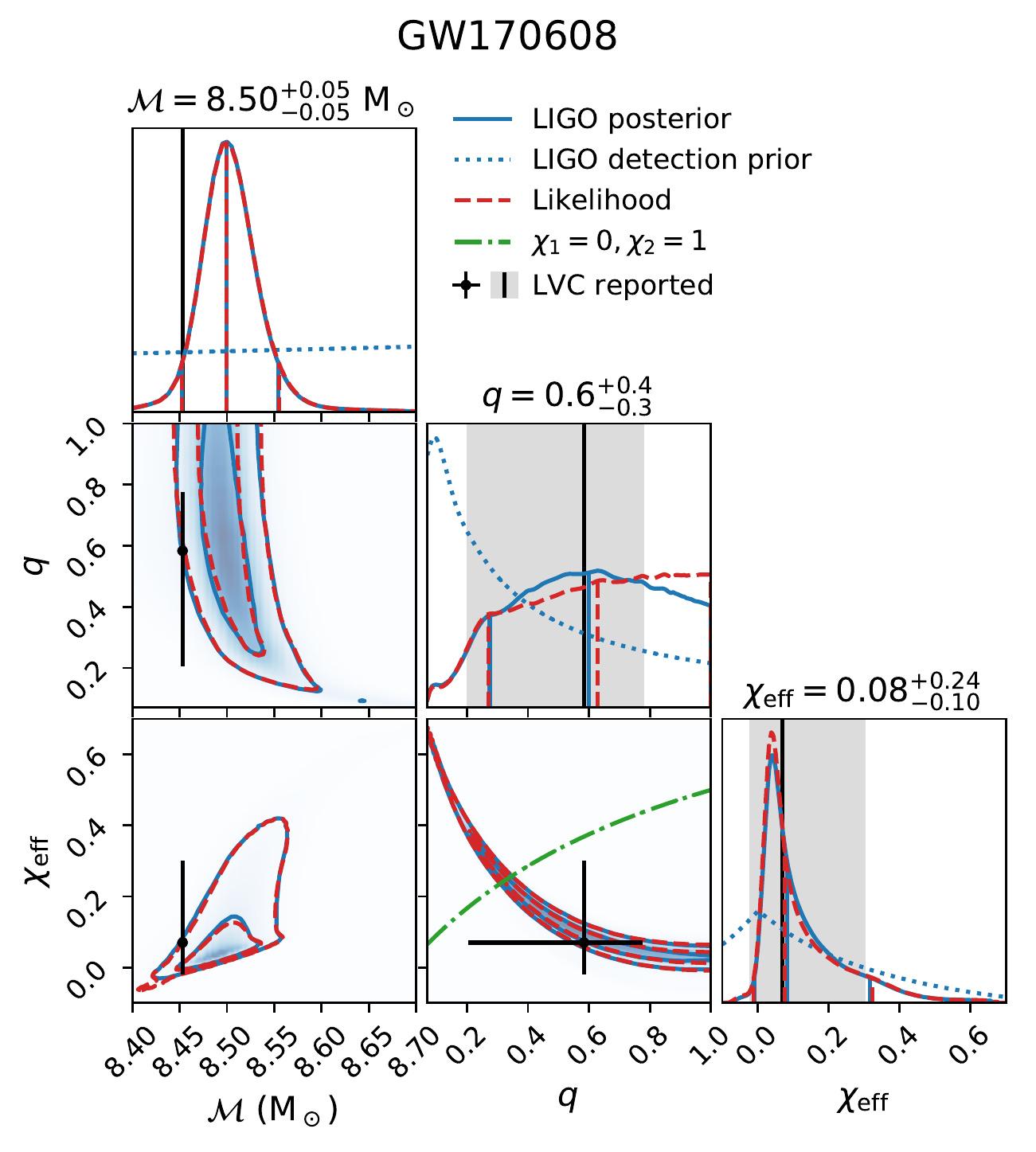}%
\includegraphics[width=.331\linewidth]{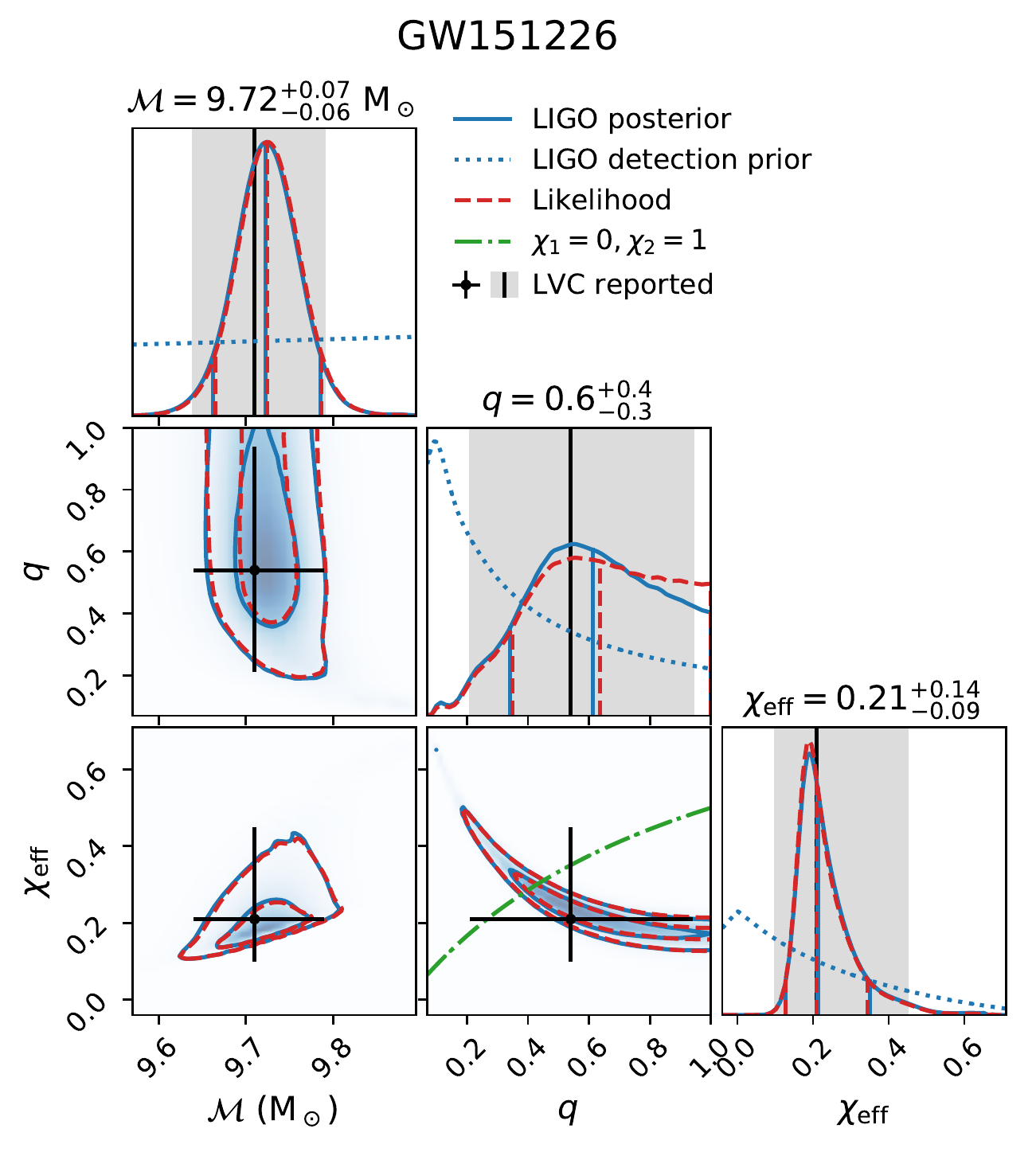}%
\includegraphics[width=.331\linewidth]{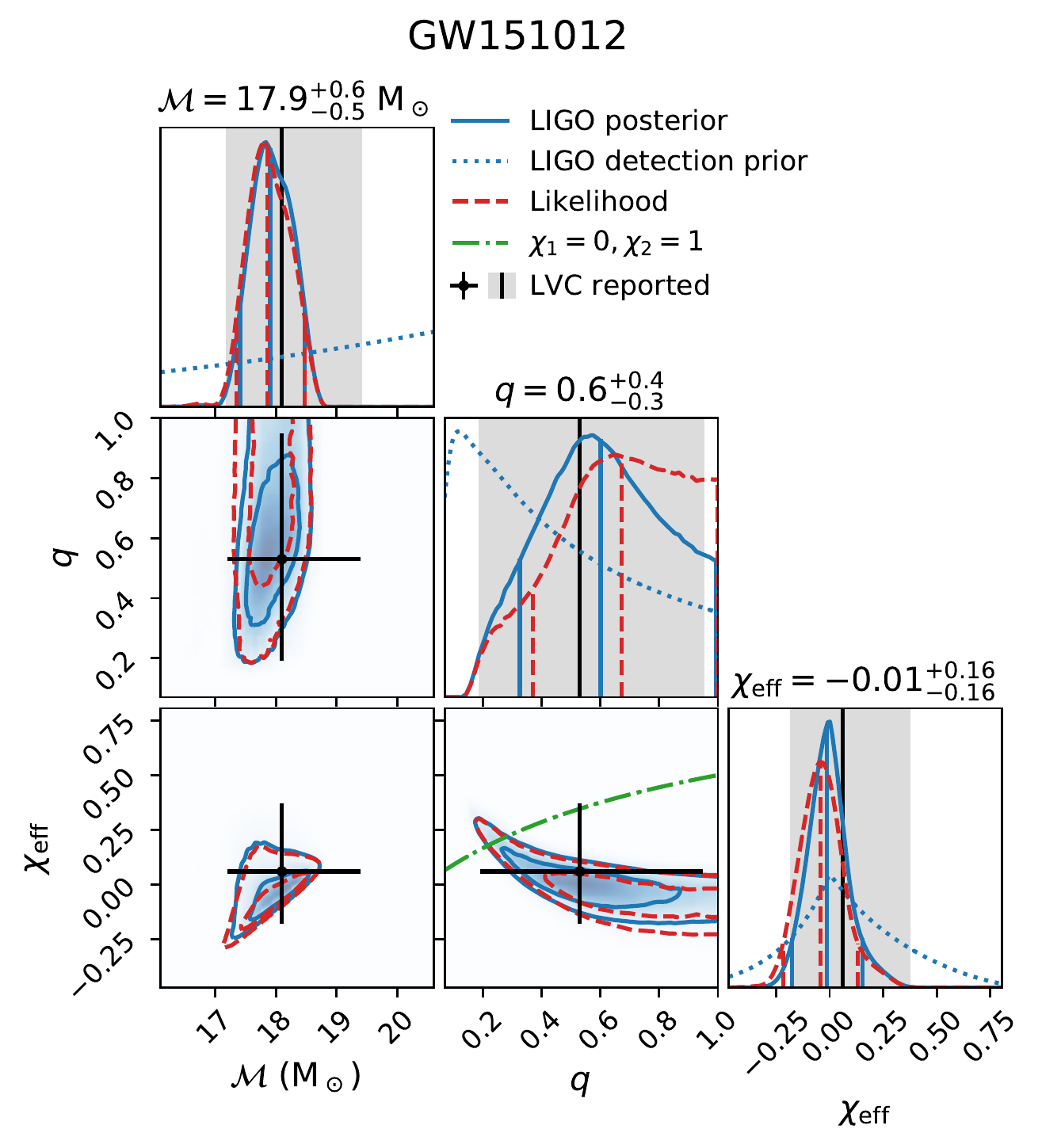}

\includegraphics[width=.331\linewidth]{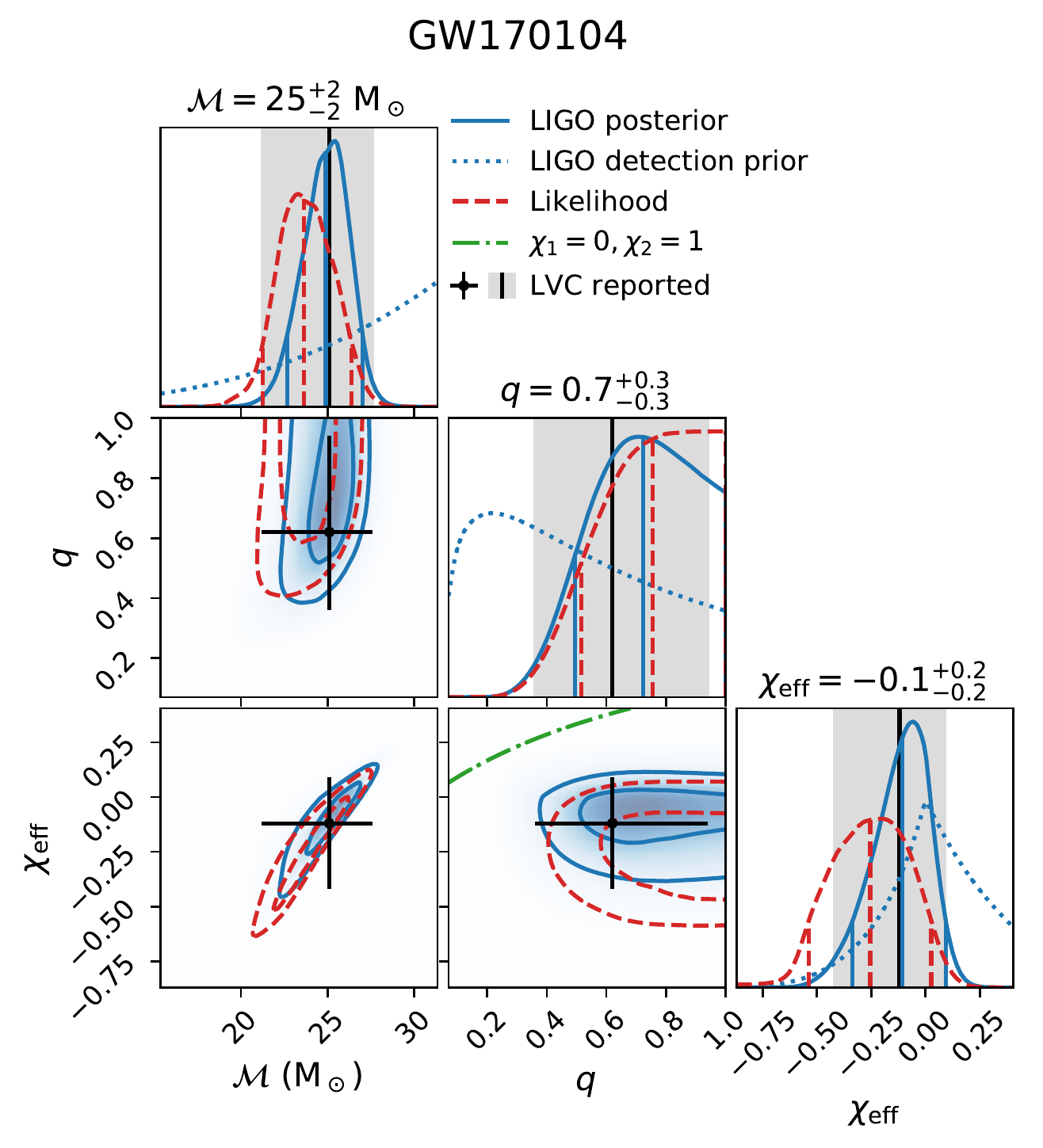}%
\includegraphics[width=.331\linewidth]{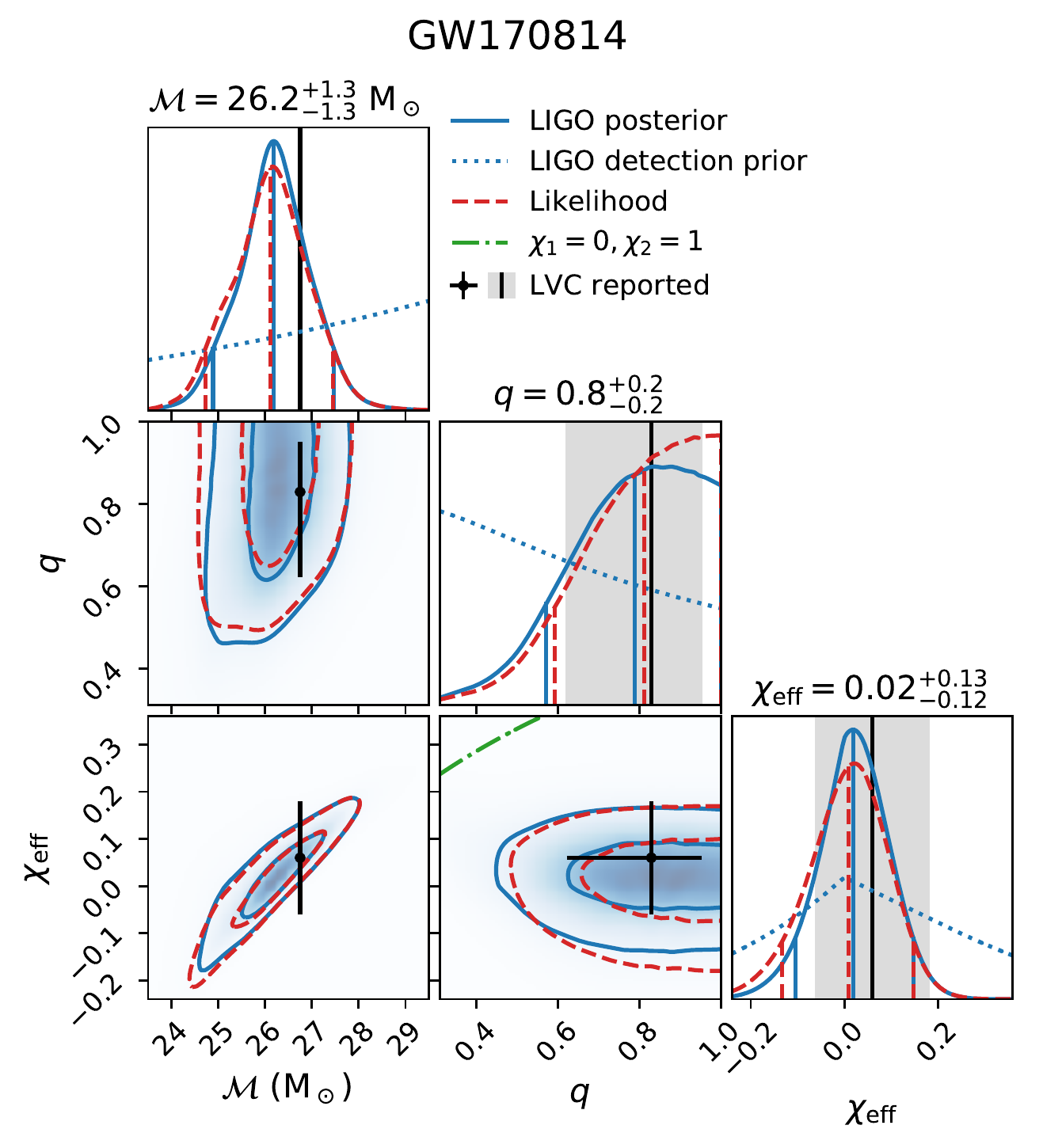}%
\includegraphics[width=.331\linewidth]{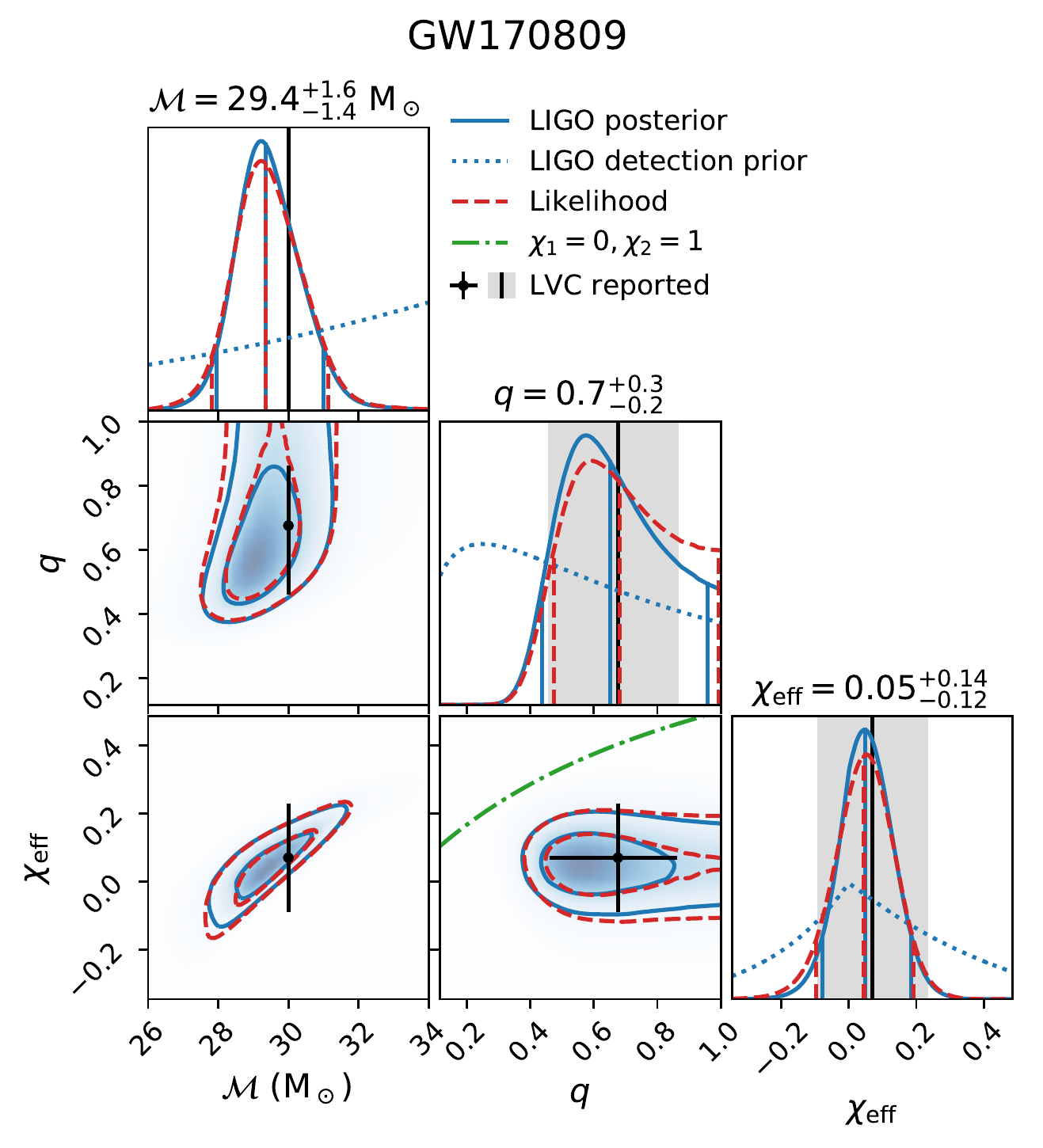}

\includegraphics[width=.331\linewidth]{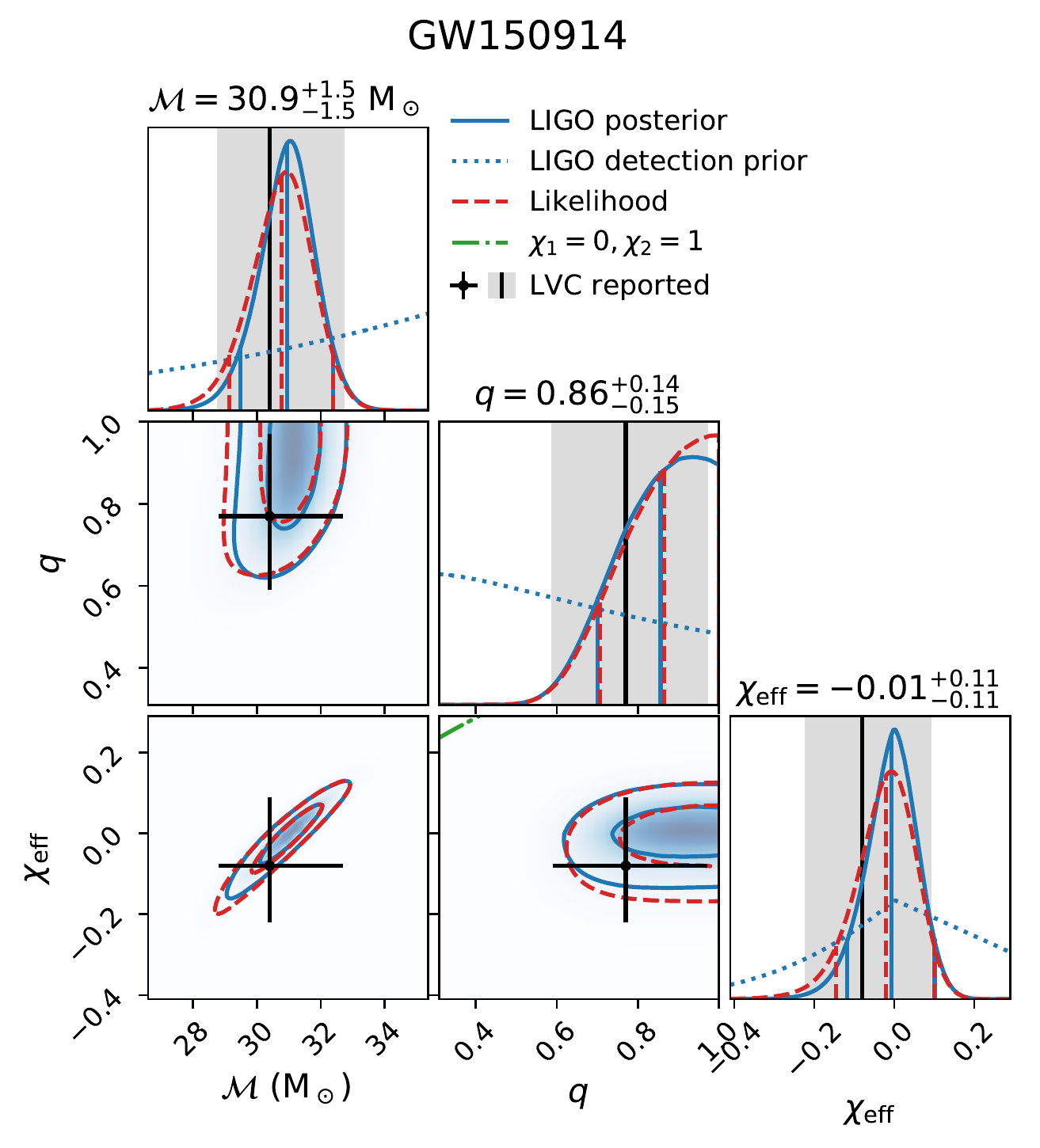}%
\includegraphics[width=.331\linewidth]{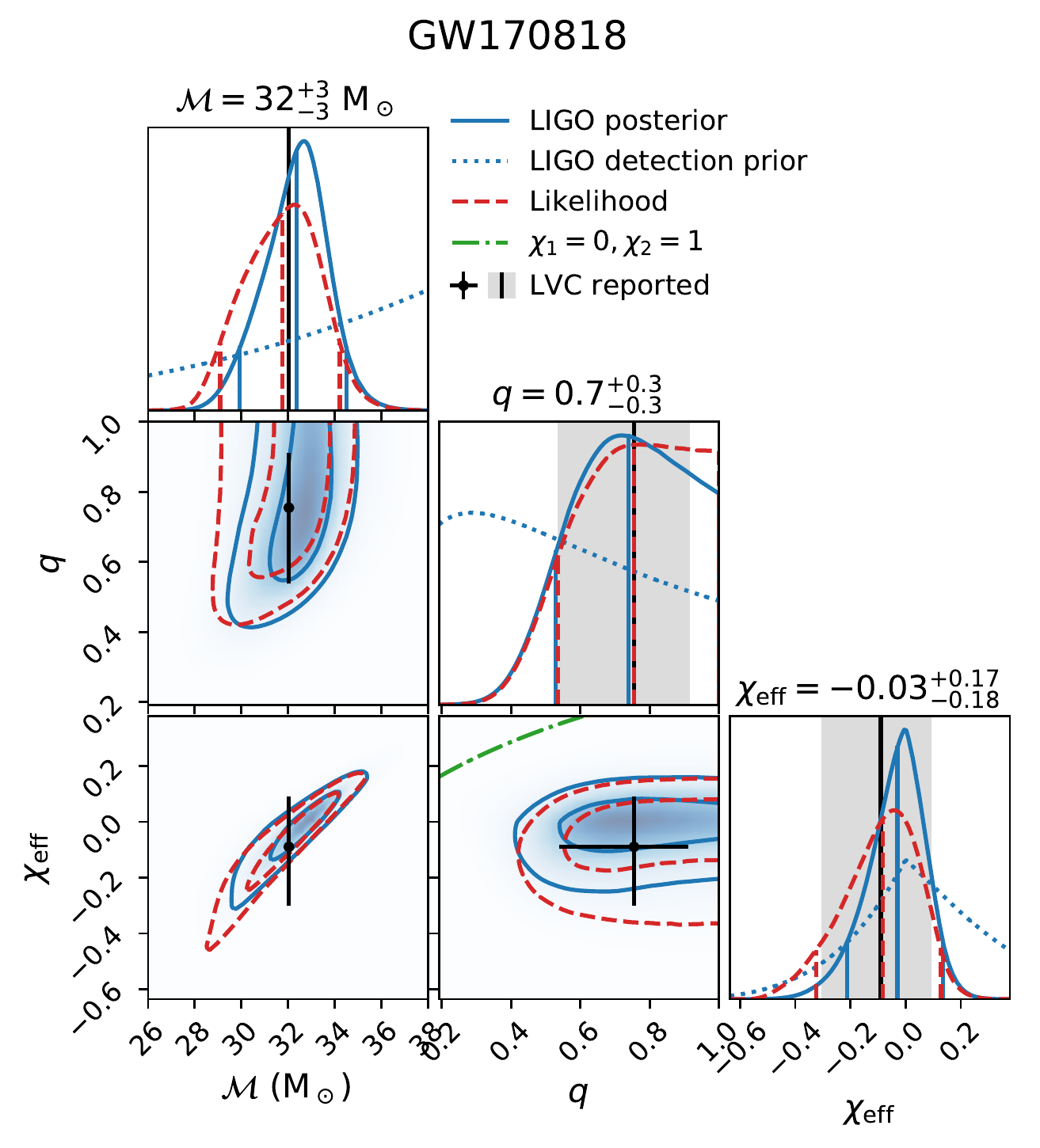}%
\includegraphics[width=.331\linewidth]{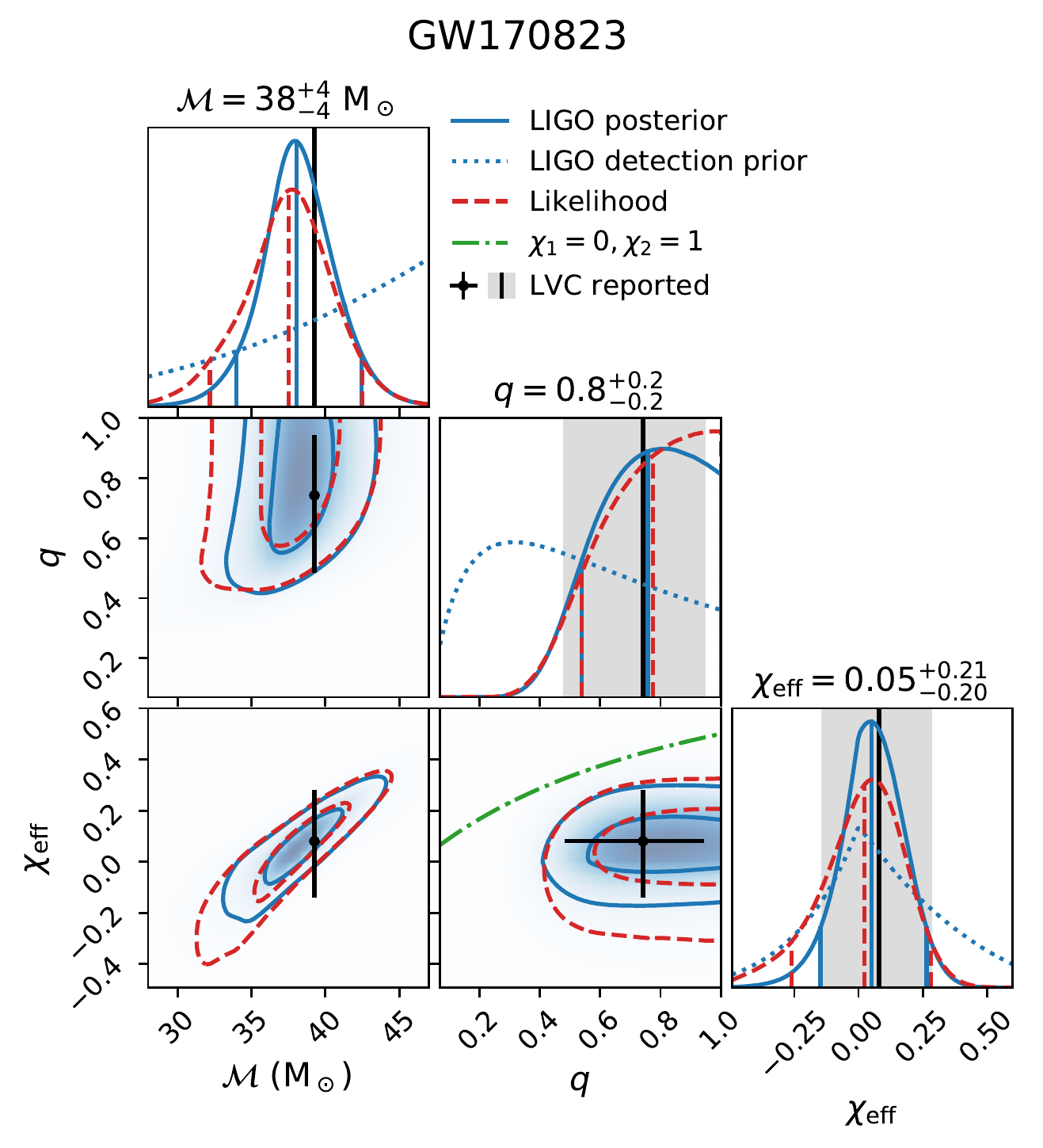}
\caption{Parameter estimation for the BBH merger events reported to date (continued in Fig.~\ref{fig:GW170729}). For each event, the plots on the diagonal show the marginalized likelihood (dashed line), detection prior (dotted) and posterior (solid) distributions for each parameter. The detection prior accounts for selection bias, and the likelihood is conditioned to detection. For each distribution, vertical lines show the median and minimal interval enclosing $90\%$. Here we use the same prior as LIGO to facilitate comparison to the reported values (black vertical line with shaded area). By ``LIGO posterior'' we mean the posterior distribution we computed using the LIGO prior.
Off-diagonal plots show the two-dimensional marginalized likelihood and posterior. Probability contours enclosing $50\%$ and $90\%$ of each distribution are shown. For the $q$--$\chieff$ plot, the case where the aligned spins of the black holes are $\chi_1=0, \chi_2=1$ is shown by a dashed-dotted line, as a proxy for what the outcome of a tidally-locked-secondary progenitor would be. The likelihood can be interpreted as the posterior distribution arising from a uniform detection prior in $\mathcal M, q, \chieff$, so it illustrates the influence of changing the prior.
The values reported by the LVC are shown by black dots with error bars.
The LVC did not report the detector-frame chirp mass for the last six events (from GW170608 on), so for those cases we show their source-frame value corrected for redshift, without an uncertainty.}
\label{fig:events}
\end{figure*}

\begin{figure}
\includegraphics[width=\linewidth]{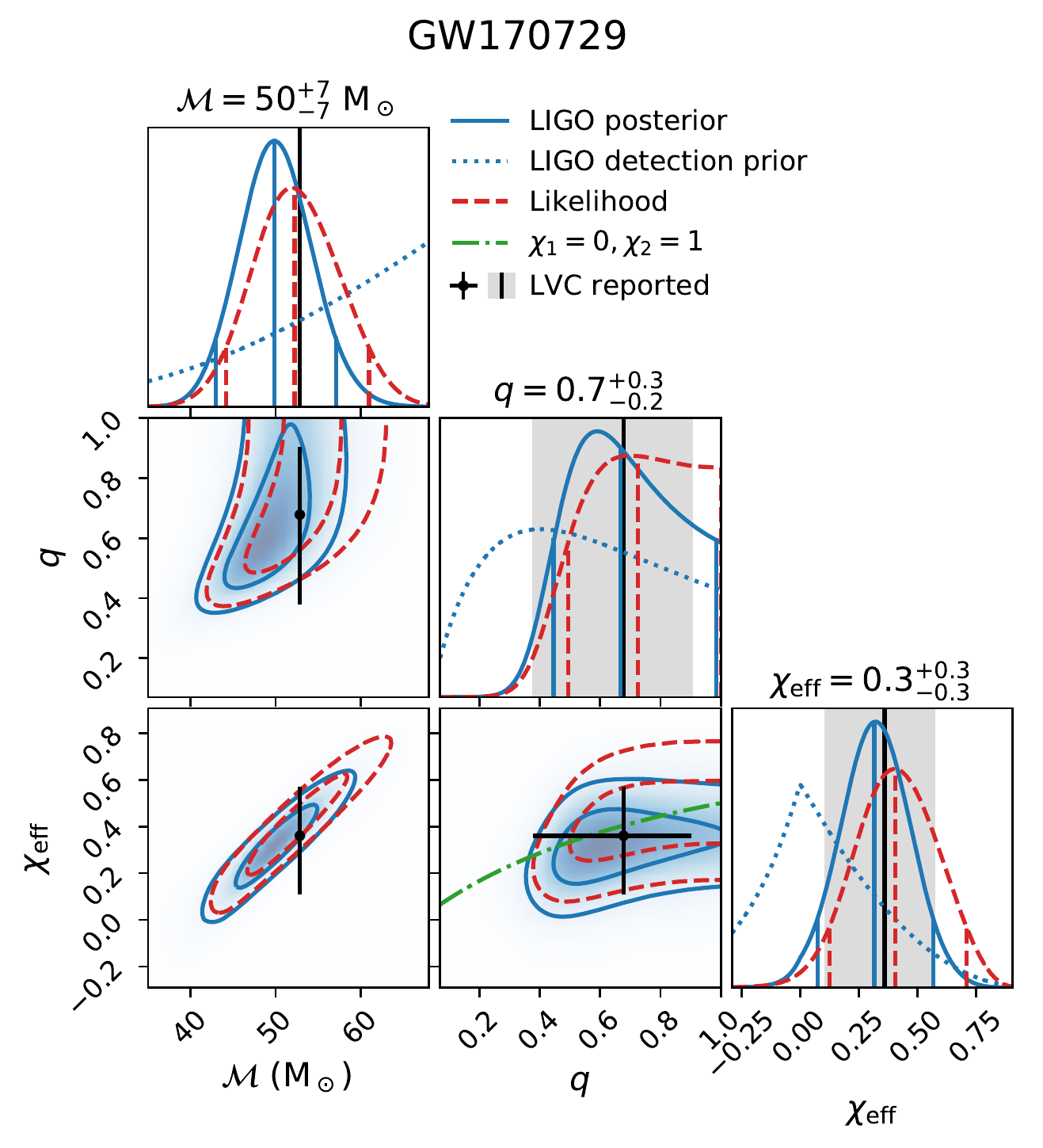}%
\caption{Continuation of Fig.~\ref{fig:events} displaying GW170729, the heaviest and highest-$\chieff$ event.}
\label{fig:GW170729}
\end{figure}

In Fig.~\ref{fig:events} we show the likelihood on the $\mathcal M, q, \chieff$ parameter space for each event, and the posterior distribution computed using the same prior as LIGO to facilitate the comparison. Our reanalysis recovers both the one- and two-dimensional marginalized posteriors accurately, which is compelling evidence that our approximations are working well and capture the degeneracies present.

The prior LIGO used for the astrophysical distribution of parameters is uniform in the individual masses, spin magnitudes and cosine of spin tilts. This induces a nonuniform prior on the variables we adopted. The transformation from $m_1, m_2$ to $\mathcal M, q$ is given by Eq.~\eqref{eq:Mchirp_q}, taking its Jacobian yields
\begin{equation} \label{eq:P_Mchirp_q_LIGO}
	P_{\rm LIGO\,prior}(\mathcal M, q) \propto \frac{1}{q} \left[ \left( 1 + \frac{1}{q}\right) (1 + q) \right]^{1/5} \mathcal M.
\end{equation}
Using Eq.~\eqref{eq:chi_eff}, we can relate the probability of $\chieff$ to the individual aligned spins $\chi_i$:
\begin{equation} \label{eq:P_chieff_given_q}
	P(\chieff \mid q) = \iint_{-1}^1 {\rm d}\chi_1 {\rm d}\chi_2 
    	\, \delta \! \left( \chieff - \frac{\chi_1 + q \chi_2}{1+q} \right)
        P(\chi_1, \chi_2).
\end{equation}
In the LIGO prior, the individual spins are uncorrelated and taken from the same distribution. Therefore, $P_{\rm LIGO\,prior}(\chi_1, \chi_2) = P_\chi(\chi_1) P_\chi(\chi_2)$, with
\begin{equation} \label{eq:P_chi}
\begin{split}
	P_\chi(\chi) &= \int_0^1 {\rm d}a \int_{-1}^1 \frac{{\rm d}\mu}{2} 
    	\delta(\chi - a \mu)\\
    &= -\frac{1}{2} \log \abs{\chi}, \qquad \abs{\chi} \leq 1.
\end{split}
\end{equation}
Using this, we can carry out the $\chi_1$ integral in Eq.~\eqref{eq:P_chieff_given_q}:
\begin{multline} \label{eq:P_chieff_given_q_LIGO}
    P_{\rm LIGO\,prior}(\chieff \mid q) \\
    = (1+q) \int_a^b {\rm d}\chi_2 
    	P_\chi \left((1+q) \chieff - q \chi_2 \right) P_\chi(\chi_2),
\end{multline}
where the integration limits are
\begin{equation}
\begin{split}
	a &= \max \left\{\frac{(1+q)\chieff -1}{q}, -1 \right\}\\
    b &= \min \left\{\frac{(1+q)\chieff +1}{q},  1 \right\}.
\end{split}
\end{equation}
In practice, we compute the integral in Eq.~\eqref{eq:P_chieff_given_q_LIGO} by quadrature (see \citet{Ng2018} for an analytical approximation). The total LIGO prior for $\mathcal M, q, \chieff$ is given by the product of Eqs.~\eqref{eq:P_Mchirp_q_LIGO} and \eqref{eq:P_chieff_given_q_LIGO}.

The LIGO detection prior shown in Fig.~\ref{fig:events} is obtained by multiplying the astrophysical prior by the sensitive volume of the detector. We show the detection prior so that any deviations of the posterior from the prior are driven by the data and not by the parameter-dependent sensitivity of the detector.

\bsp	
\label{lastpage}
\end{document}